\documentclass[article]{aa}  
\usepackage{graphicx}  
\usepackage{txfonts}  
\usepackage{natbib}
\usepackage{lscape}

\bibpunct{(}{)}{;}{a}{}{,} 
  
\usepackage{color}
\usepackage[normalem]{ulem}

\begin{document}  

\title{Chromospheric activity and photospheric variation of $\alpha$ Ori during the great dimming event in 2020}  
  
\author{M. Mittag\inst{1} \and K.-P. Schr\"oder\inst{2} \and V. Perdelwitz\inst{3,1} \and D. Jack\inst{2} \and J.H.M.M. Schmitt\inst{1}}
  
\institute{Hamburger Sternwarte, Universit\"at Hamburg, Gojenbergsweg 112,   
21029 Hamburg, Germany\\  
\email{mmittag@hs.uni-hamburg.de}  
\and   
Departamento de Astronomia, Universidad de Guanajuato, Mexico
\and
Kimmel fellow, Department of Earth and Planetary Science, Weizmann Institute of Science, Rehovot, Israel
}
  
\date{Received \dots; accepted \dots}  
  
\abstract  
{}
{The so-called great dimming event of $\alpha$ Ori in late 2019 and early 2020 sparked our interest in the behaviour of
  chromospheric activity during this period. $\alpha$ Ori  was already part of the 
  long-term monitoring program of our TIGRE telescope to study the stellar activity of giant stars, and therefore regular measurements
  of $\alpha$ Ori have been taken since 2013.}  
{In the context of this study, we determined the TIGRE S-index values and, using a set of
  calibration stars, converted these to the Mount Wilson S-index scale, which allows us
  to combine our TIGRE activity measurements with the S$_{MWO}$ values taken
  during the landmark Mount Wilson program some decades earlier and to compare
  that extended time series with the visual and V magnitude photometric data 
  from the AAVSO database.
  In addition, we determined the absolute and normalised excess flux of 
  the \ion{Ca}{II} H\&K lines.
  To understand the activity in absolute terms, we also assessed the changes in effective 
  temperature using the TiO bands covered by our TIGRE spectra.} 
{We find a clear drop in effective temperature by about 80 K between 
  November 2019 and February 2020, which coincides with the minimum of visual 
  brightness. In addition, the effective (luminous) photospheric area of 
  $\alpha$ Ori also shrank.  This might be related to a temporary
  synchronisation of several large convective cells in cooling and
  sinking down. During the same period, the S-index increased significantly, 
  yet this is a mere contrast effect, because the normalised 
  excess flux of the \ion{Ca}{II} H\&K lines did not change significantly.
  However, the latter dropped immediately after this episode. Comparing the combined
  S$_{MWO}$ values and visual magnitude time series, we find a similar increase
  in the S-index during another noticeable decrease in the visual magnitude of 
  $\alpha$ Ori which took place in 1984 and 1985. These two episodes of dimming therefore seem to share a common nature.
  To probe the dynamics of the upper photosphere, we further analysed the closely 
  neighbouring lines of \ion{V}{I} and \ion{Fe}{I} at 6251.82 \AA~and 6251.56 
  \AA~respectively. Remarkably, their core distance varies, and once converted to radial
  velocity, shows a relation with the great dimming event, as well as with the 
  consecutive, weaker dimming episode in the observing season of 2020 and 2021. This 
  type of variation could be caused by rising and sinking cool plumes as a temporary spill-over of convection on $\alpha$ Ori.}  
{As the effective temperature of $\alpha$ Ori  is variable, the S-index, computed relative to a
  near-ultraviolet (NUV) continuum, is only of restricted
  use for any monitoring study of the chromospheric activity of $\alpha$ Ori. 
  It is therefore important to consider the effective temperature variability 
  and derive the normalised \ion{Ca}{II} H\&K flux to study the chromospheric
  long-term changes in absolute terms. In fact, the \ion{Ca}{II} H\&K 
  normalised excess flux time series shows that the chromospheric emission of 
  $\alpha$ Ori  did 
  not change significantly between November 2019 and February 2020,
  but then beyond the great dimming minimum it does vary. Hence, this delay 
  of the chromospheric reaction suggests  that the cause for the  great dimming
  is located in the photosphere.   An investigation of the long-term 
  spectroscopic and photometric time series of $\alpha$ Ori suggests that
  the great dimming in 2019 and 2020 does not appear to be a unique 
  phenomenon,  but rather that such dimmings do occur more frequently, which motivates further
  monitoring of $\alpha$ Ori with facilities such as TIGRE.}

\titlerunning{Chromospheric activity and of photospheric variation $\alpha$ Ori}  
  
\maketitle  
   
\section{Introduction}

The red supergiant $\alpha$ Ori (HD~39801, Betelgeuse) is one of the brightest stars in 
the sky and showed an unusual decrease in its brightness in the winter between 2019 and 2020.
Normally, the brightness of $\alpha$ Ori typically varies between $\sim$0.2 and $\sim$1 mag 
in the Johnson V band, as demonstrated by a light curve available in the AAVSO
(American Association of Variable Star Observers)
database\footnote{\citet{Kafka},https://www.aavso.org}.
However, in February 2020, the brightness of $\alpha$ Ori decreased 
extraordinarily to a mere V$\sim$1.6, an event that was consequently dubbed
the `great dimming event'. In December 2019, \citet{Guinan2019ATel13341} 
raised public awareness of this event, and an ESO press release\footnote{https://www.eso.org/public/news/eso2003/}
showed VLT/SPHERE (Very Large Telescope)/(Spectro-Polarimetric High-contrast Exoplanet REsearch instrument)
observations of $\alpha$ Ori by Montarg\`{e}s, comparing its state in January 2019 with that in 
December 2019; this VLT/SPHERE image confirmed the significant darkening of Betelgeuse
in comparison to an image taken in January 2019.

As to the cause of the unusual decrease in brightness, two main hypotheses were put
forward: the first interprets the cause of the dimming to be a decrease in
effective temperature, that is, a photospheric effect, while the alternative 
hypothesis argues that the photospheric emission had been 
shielded by material from some ejection or increased mass-loss event, thus located 
further out and not directly related to the photosphere.

\citet{Levesque2020ApJ} studied low-resolution spectra of $\alpha$ Ori; while finding
a small drop in effective temperature, the authors argue that this decrease is too small to 
explain the great dimming and suggest instead a mass ejection as an explanation for the observed decrease in brightness.
This scenario is supported by \citet{Dupree2020ApJ}, who analysed the mid-ultraviolet (MUV) flux in 2400-2700 \AA~
and \ion{Mg}{II} h\&k lines with spectra taken with HST/STIS (Hubble Space Telescope)/(Space Telescope Imaging Spectrograph),
and found that both fluxes, as measured in September 2019, October 2019, and November 2019, were
larger than in February 2020 at the maximum of the great dimming event. Therefore, \citet{Dupree2020ApJ}
argue that a great mass ejection occurred in October 2019 leading to the formation of a dust cloud
in the following weeks, which eventually darkened $\alpha$ Ori. 
Dust dimming of $\alpha$ Ori is also supported by the work of \citet{Montarges2021Natur}, who
analysed VLT/SPHERE images taken in January 2019, December 2019, January 2020, and March 2020. 

On the other hand, the dust cloud theory was disputed by \citet{Dharmawardena2020ApJ}, 
who investigated submillimetre (submm) data taken with the James Clerk Maxwell Telescope and 
Atacama Pathfinder Experiment.  While \citet{Dharmawardena2020ApJ} also
found $\alpha$ Ori to be about 20\% fainter in the submm range, their radiative modelling 
suggests that the dimming event was 
caused by some photospheric  change. More specifically, a change in effective 
temperature during the great dimming was also demonstrated by \citet{Harper2020ApJ}, 
who present a TiO band light curve; based on these observations, the derived 
effective temperature time series shows a significant drop, which coincides 
well with the great dimming event as observed in the V band. 

In this paper we present our $\alpha$ Ori observations obtained with the 
1.2m TIGRE telescope (for more details, see \citet{Schmitt2014AN335787S,Gonzalez-Perez_2022} and below), 
an instrument designed primarily for the study of stellar chromospheric
activity. However, before we present our respective results in more detail, we need
to address the changes in effective temperature, because these imply physical 
changes of the photosphere leading to changes in the spectral fluxes, which are used 
as a reference in many activity indicators. 
Consequently, the drop in temperature 
around February 2020 has a direct impact on the interpretation of the observed
activity indicators, such as the
calibrated Mount Wilson S-index, as well as the derivation of the absolute 
\ion{Ca}{II} H\&K line flux of $\alpha$ Ori. Thereafter, we present a combined 
time series of the Mount Wilson S-index and compare its behaviour to photometric data. 

Furthermore, we derive the absolute and excess \ion{Ca}{II} H\&K line flux, which
show a remarkably different relation with the \mbox{V-band} light curve. Also, we present 
the significant line depth and radial-velocity (RV) variations of the unequal 
neighbouring line pair of \ion{V}{I} 6251.82 \AA~and \ion{Fe}{I} 6252.56 \AA.
These lines form in different depths of the photosphere and are indicative of 
a height-dependent dynamical behaviour of the photosphere apparently
associated with the great dimming event. Finally, we summarise our results and
present our conclusion about the possible physical causes for this dimming event.

\section{TIGRE spectroscopic monitoring and data reduction}

TIGRE (Telescopio Internacional de Guanajuato Rob\'otico Espectrosc\'opico)
is a fully robotic spectroscopic telescope with an aperture of 1.2~m, located at 
the La Luz Observatory of the University of Guanajuato, Mexico,  designed for long 
term monitoring programs.  The two-spectral-channel, fibre-fed \'Echelle spectrograph HEROS
(Heidelberg Extended Range Optical Spectrograph) is located in a thermally and mechanically isolated room. The 
spectral resolving power of the HEROS spectrograph is R$\approx$20,000, which is relatively uniform 
over a wavelength range from 3800 \AA~to 8800 \AA~with a minor gap around 5800~\AA~
between the two spectral channels caused by the dichroic beam splitter; the TIGRE facility
and the realisation of its robotic operation is described in more detail by \citet{Schmitt2014AN335787S} and \citet{Gonzalez-Perez_2022}. 

The TIGRE/HEROS spectra used here were reduced with the TIGRE automatic standard 
reduction pipeline v3.1 written in IDL based on the reduction package 
REDUCE \citep{REDUCE2002A&A385.1095P}. This latter pipeline includes all required reduction steps for 
an \'Echelle spectrum; a detailed description of the first version of the TIGRE reduction
pipeline is given by \citet{mittag2010}, and additional information can be found
in \citet{Hempelmann2016} and \citet{Mittag2016A&A}.

The TIGRE observing plan includes a monitoring program of the chromospheric activity  
of giant stars, providing a study of the variations of the \ion{Ca}{II} H\&K lines. 
One target of this program is the red supergiant $\alpha$ Ori. Until November 2019, 
with the goal of studying long-term variability, $\alpha$ Ori was for the most part observed with a monthly 
cadence; However, in December 2019,  we started covering $\alpha$ Ori more
frequently to better monitor the stellar activity during the great dimming event.
In Table \ref{obs_alphaOri}, we list the number of observations taken in each 
observing season and the range of the signal-to-noise ratio (S/N) at 4000 \AA~per season. 
\begin{table}[!t]    
\caption{$\alpha$ Ori observations by observing season (September-April)}    
\label{obs_alphaOri}    
\begin{center}    
\begin{small}
\setlength{\tabcolsep}{5pt}
\begin{tabular}{c c c}
\hline    
\hline
Epoch & No. & $\langle \rm{S/N} \rangle$ range \\
\hline
\noalign{\smallskip}
2013/14 & 15 & 39-98 \\
2014/15 & 4  & 180-272 \\
2015/16 & 3  & 173-207 \\
2016/17 & 1  & 311 \\
2017/18 & 2  & 526-554 \\
2018/19 & 2  & 407-448 \\
2019/20 & 15 & 103-554 \\
2020/21 & 32 & 22-399 \\
\hline
\end{tabular}
\end{small}
\end{center}    
\end{table}

\section{Assessment of long-term effective temperature variations}
\label{teff_est}

Before discussing the chromospheric activity of $\alpha$ Ori,
we try to quantify the possible variations of the effective temperature during the great dimming event 
in order to consider this in our quantitative analysis of the chromospheric 
activity at this point in time. 
To begin with, we used the publicly available photometric data from the AAVSO database
as a reference to the timing of the great dimming event. Various colour indices provide 
an indicator of the effective temperature. However, since their variation could in
principle be caused partly by extinction from dust, we also use TiO bands  here
to estimate the effective temperature from our TIGRE/HEROS spectra
by comparing the TiO bands of each TIGRE/HEROS spectrum
with best-matching PHOENIX atmospheric model spectra.

\subsection{Time series of photometric data and colour indices}

\begin{figure}
\centering
\includegraphics[scale=0.38]{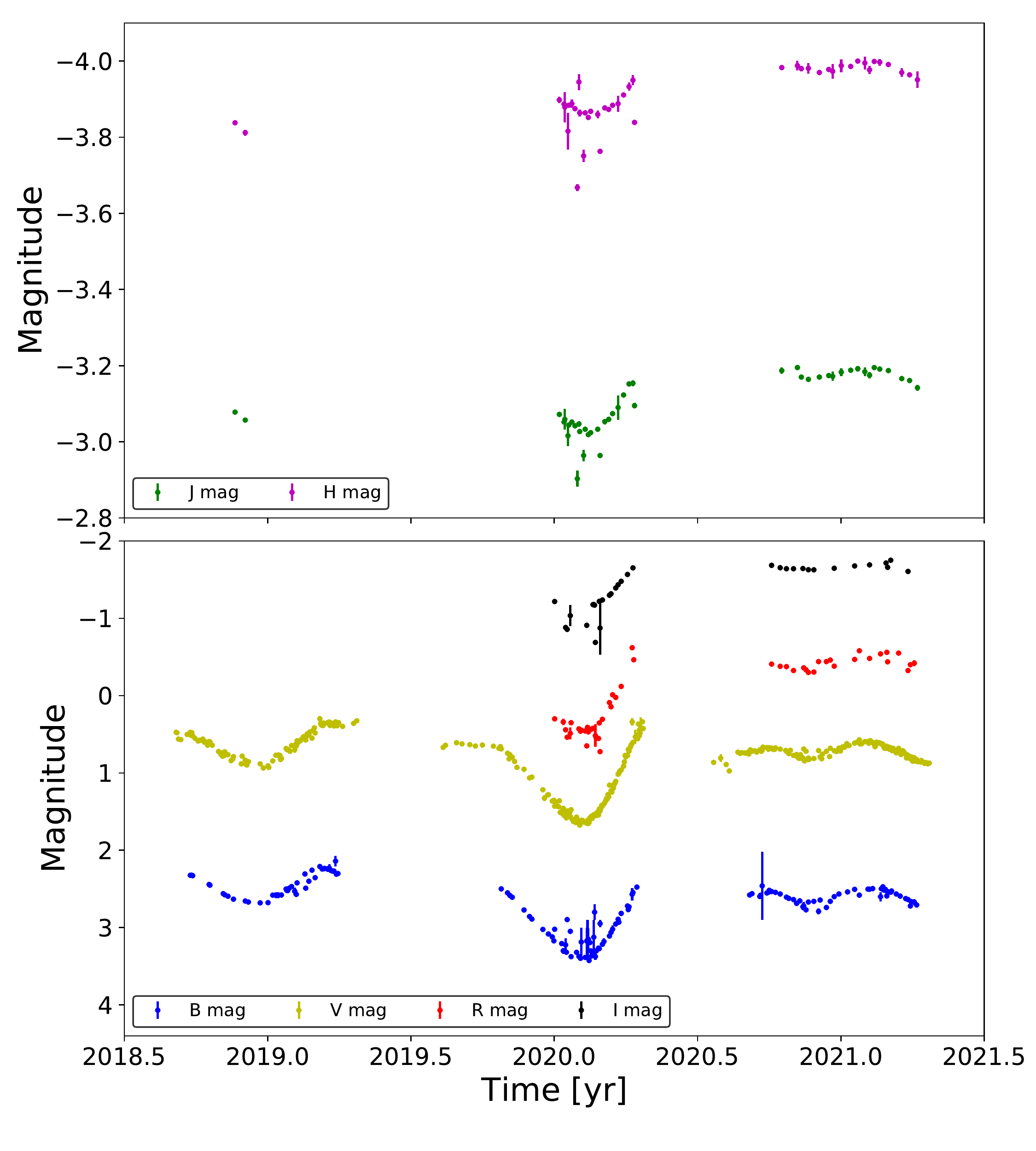}
\caption{Light curves of the B, V, R, I, J, and H bands, beginning with 2018.5.
        The upper panel depicts the J and H magnitudes, and the lower panel shows the 
        B, V, R, and I bands.}
\label{colour_time_series}
\end{figure}

\begin{figure}
\centering
\includegraphics[scale=0.38]{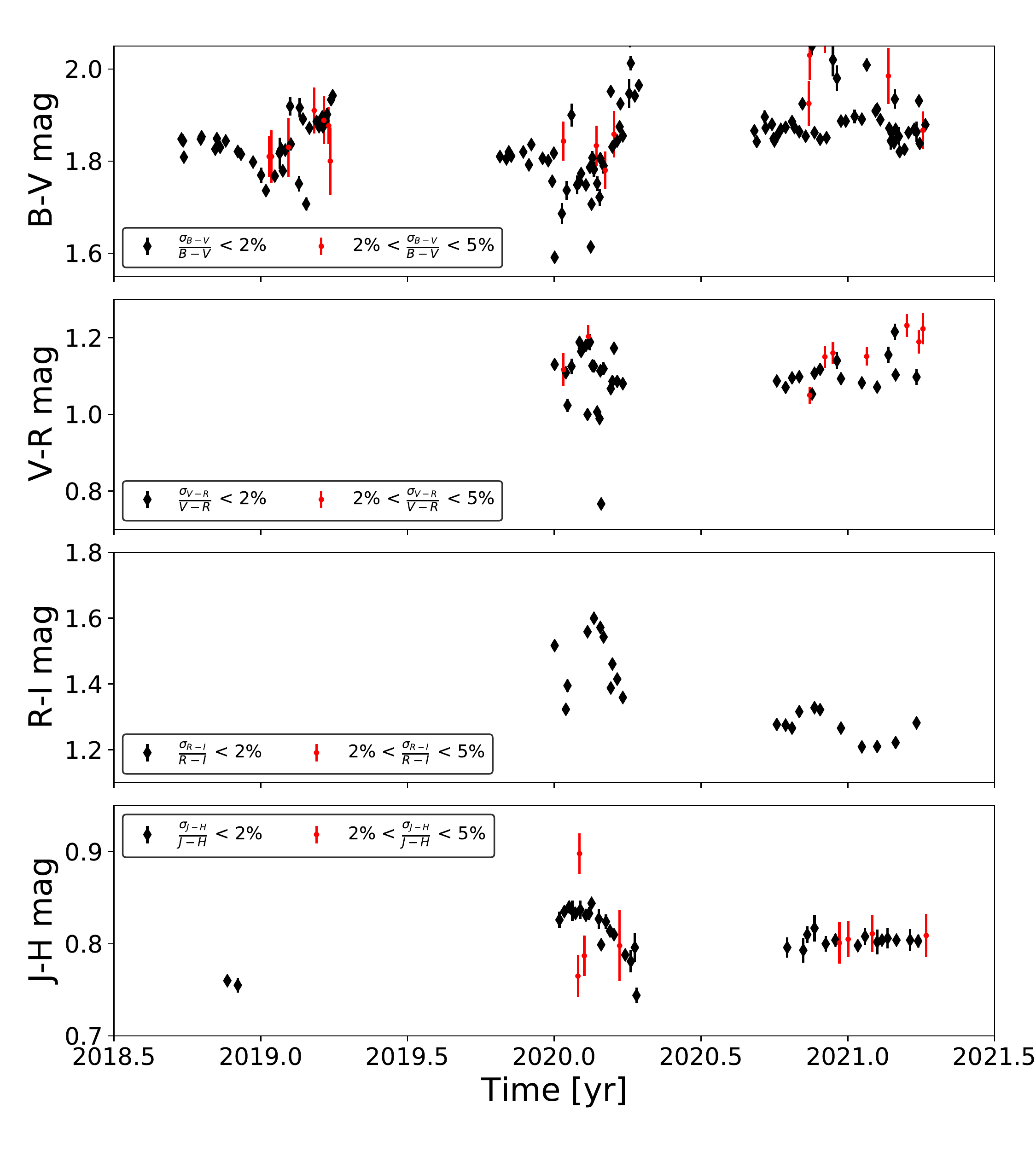}
\caption{Colour index time series. The top panel shows the $B-V$ values, 
        the second displays the $V-R$ values, the third the $R-I$ values,
         and the lowest panel depicts the $H-J$ values. Black dots represent
         values with a relative error of lower than 2$\%$, whereas red data points those lower than 5$\%$.}
\label{colour_index}
\end{figure}

Starting with epoch 2018.5, in Fig. \ref{colour_time_series} we show the daily brightness 
in the B, V, R, I, J, and H bands, as listed in the AAVSO database. In all bands, the light curves
show a variation apparently caused by a 
pulsation cycle, its amplitude being smaller in the J and H bands.
Next,  we computed the $B-V$, $V-R$, $R-I,$ and $J-H$ colour indices for every day with
available data; to ensure that any variation is real, we only provide values with a relative 
error of lower than 5\% of the respective index. These values are shown in Fig. \ref{colour_index},
which demonstrates
that variations occur in all four colour indices; the times and values of these changes are 
consistent with the effective temperature variation shown by \citet{Harper2020ApJ}.
The colour indices $R-I$ and $J-H$ increase during the great dimming event, which
shows a minimum in the light curve in each band. 
Both indices show the behaviour expected, that is, a decrease in the effective
temperature, which manifests as an increase in these two indices; see Sect. \ref{phx_colour}.
However, the colour index $V-R$  shows strong scatter 
during the minimum of the light curves, without any clear trend,  and the $B-V$ colour 
index actually decreases during the great dimming event. 

While over a large temperature range in cool stars the B-V index increases with 
decreasing effective temperature, this is not the case here ---the 
coolest stars form an exception. In fact, $B-V$ values calculated from PHOENIX 
spectra of an effective temperature of less than 3600 K show the same behaviour.
Above 3600 K, with rising effective temperature, the $B-V$ values also decrease 
from their maximum value there, but very slowly  at first. 
The $B-V$ values based on PHOENIX spectra \citep{Husser2013A&A} are shown
in Fig. \ref{phx_colour_indices} and further explanation of their derivation is given in 
Sect. \ref{phx_colour}. 
This inverted $B-V$ behaviour was also found in the super giant star
HD~156014 ($\alpha$ Her A). This object has a $B-V$ colour of only 1.164 mag \citep{HIPPARCOS1997ESA},
but an effective temperature as low as 3271 K 
(taken from the PASTEL catalogue, Version 2020-01-30; \citep{Soubiran2016}). 
Interestingly, a similar trend inversion was shown by \citet{lancon2007A&A} 
for the $H-K$ index, calculated from PHOENIX spectra for red supergiants and giants.

\subsection{Effective temperature assessment based on TiO bands}
\label{teff_tio_est}

\begin{figure}
\centering
\includegraphics[scale=0.36]{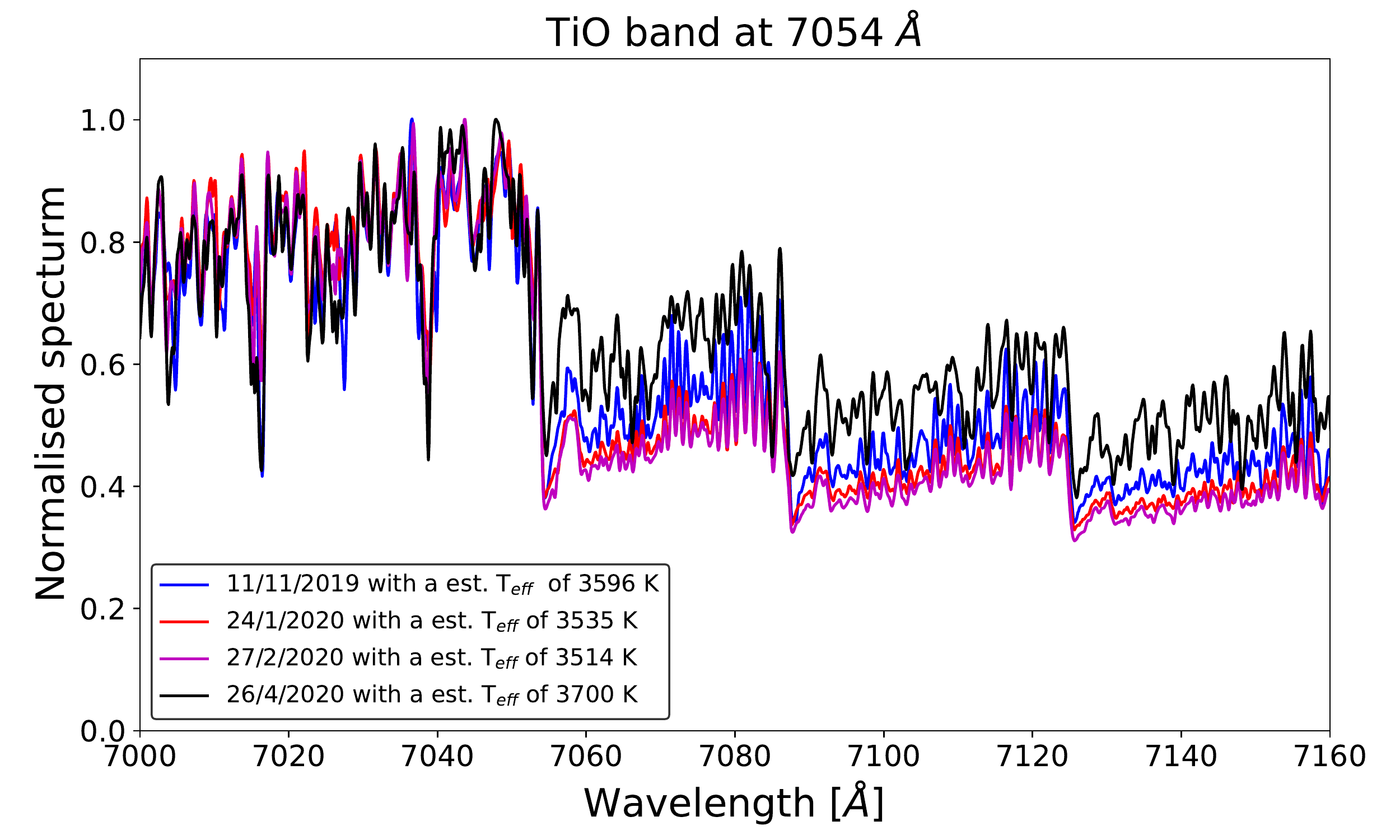}
\caption{TiO band at 7054 \AA~taken on four different dates: 11 November 2019 (blue), 
        24 January 2020 (red), 27 February 2020 (magenta), and 26 April 2020 (black).}
\label{tio_band}
\end{figure}

To perform a quantitative assessment of any effective temperature variations, we now 
use the TiO bands; these bands become stronger in cooler stars and thereby provide
a good effective temperature indicator in their own right. 
Examining the TiO band at 7054 \AA,   in Fig. \ref{tio_band}
we compare   
four TIGRE spectra  taken before, during, and after the great dimming 
event and indeed find a decrease in their normalised fluxes during 
the great dimming; this finding indicates that the effective temperature dropped during 
the great dimming event.

To quantify the effective temperature based on the TiO bands strength, we compared four normalised 
TiO bands at 4954 \AA, 5450 \AA,
6154 \AA,~and 7054 \AA~in the TIGRE spectra with normalised PHOENIX spectra of the 
G\"ottingen University database created by \citet{Husser2013A&A},
in a temperature range from 3300~K to 3900~K (with a step size of 100 K, a log g of 0.0,
and solar metallicity). For this comparison, the rotational line-broadening (v$\cdot$sin(i)) and a spectral 
resolving power of R$\approx$20000 are considered in the PHOENIX spectrum. 

\begin{figure}
\centering
\includegraphics[scale=0.36]{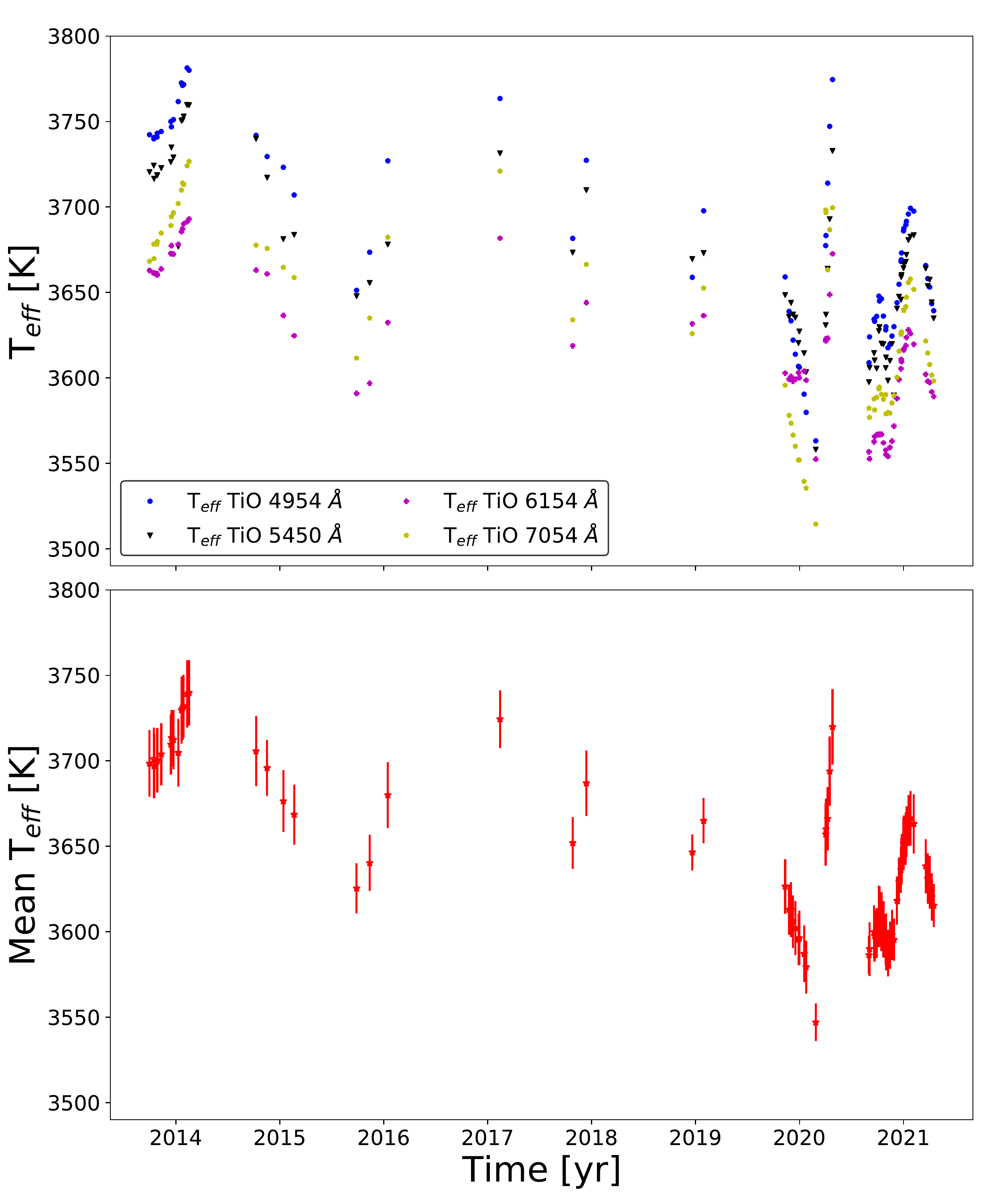}
\caption{Effective temperature time series obtained from TIGRE/HEROS spectra:
  Upper panel: Effective temperatures obtained from four TiO bands. 
  Lower panel: Respective averaged effective temperatures.}
\label{tio_teff}
\end{figure}

To estimate the differences between observed spectra and PHOENIX spectra, 
and to find the best matches, we computed the $\chi^{2}$ values of the residuals
for this effective temperature range  of the PHOENIX models. 
 The effective temperatures determined in this way 
for each band and each TIGRE spectrum 
are shown in the upper panel of Fig. \ref{tio_teff}. Ideally, these different 
temperatures should agree, but systematic differences illustrate the remaining 
shortcomings of the models in extreme non-local thermodynamic equilibrium (non-LTE) conditions
under very low gravity and in very extended and possibly non-spherically symmetric atmospheres; 
inaccuracies in the opacities may also influence the agreement here. In order to nevertheless
be able to derive a single effective temperature 
for each spectrum, the results for the four TiO bands were simply 
averaged, and the standard error of the mean is used as uncertainty for the resulting 
value of the effective temperature. In this way, we may at least have the
best-possible account of the temperature variations on a relative scale; these 
results are shown in the lower panel of Fig. \ref{tio_teff}. 

Figure \ref{tio_teff} clearly suggests that the effective temperature indeed 
dropped during the great dimming, with a minimum at epoch 2020.1, and a subsequent increase.
For the observation of 11 November 2019, at the onset of the great dimming, we derive 
an effective temperature of 3627$\pm$16 K, well below the level of around 3700~K observed in 
previous years. The TiO-band-derived effective temperature then decreases further  
to its minimal value of 3547$\pm$11 K in the TIGRE spectrum of 27 February 2020. This 
is then followed by a clear increase in temperature, with a small
temperature drop in November 2020. For completeness, the mean effective temperature 
for all TiO band measurements was computed. We obtained a mean effective temperature 
of 3649$\pm$6 K with a standard deviation of 48 K for $\alpha$ Ori over the past 6 years.

Our work shows fair agreement with the work of \citet[Fig.1]{Harper2020ApJ}, which 
is based on TiO band photometry, simply by visual comparison of the respective
effective temperature time series. 
\citet{Harper2020ApJ} mention two effective temperature values,
one for September 2019 of 3645$\pm$15 K, and another for February 2020 of 3520$\pm$25 K.
The latter value compares directly with our measurement of 27 February 2020, which is
in the same time range of the seven-day bins used by \citet{Harper2020ApJ}.
This comparison shows that our derived effective temperatures agree to within
the errors with the values obtained by \citet{Harper2020ApJ}. The same conclusion can be drawn 
from a comparison of the September value of \citet{Harper2020ApJ}
with our November observation. 

In addition, we note that, if there were
contributions from regions of different temperature, the flux weighting would always
favour the hotter regions, in the sense that a cool patch would have a much smaller 
impact on the overall value of the effective temperature than suggested by its area fraction. 
Consequently, the non-uniformity of the dynamic photosphere of $\alpha$ Ori
may imply much larger local temperature variations than the ones found here on
a global scale.

Based on the evidence given by the colour indices and our effective temperature values 
derived from the TiO bands, we conclude that the effective temperature
of $\alpha$ Ori is indeed variable, as is the spectral energy distribution of $\alpha$ Ori as a consequence. 
This is a particularly critical point for UV and extreme-UV (EUV)
continuum fluxes far down the short-wavelength tail end of the spectral energy
distribution of this very cool supergiant and therefore this flux depends on the effective
temperature by a very large power ($o(10)$). Consequently, this result is very 
important for the quantification of the chromospheric activity, which is presented 
below.

\section{Chromospheric activity}

A well-known and commonly used indicator for stellar chromospheric activity is the
line emission of the 
\ion{Ca}{II} H\&K doublet line at 3968.47 \AA~and 3933.66 \AA, which is characterised by the
so-called S-index.
During the great dimming event of $\alpha$ Ori, 
our TIGRE observations show strong changes in the observed S-index, which we
present in this section. These measurements, combined with historical Mount Wilson S-measures,
provide a long-term time series 
for $\alpha$ Ori, which yields an interesting comparison to the available photometric data. 
Finally, we discuss the significant temperature impact on the S-values and present an 
analysis of the \ion{Ca}{II} H\&K fluxes in absolute terms.

\subsection{\ion{Ca}{II} K variation during the great dimming}

Upon visual inspection of the \ion{Ca}{II} K line spectra taken by TIGRE during the 
great dimming event we see a strong increase in the line core relative to the surrounding
photospheric profile. For a closer comparison, all spectra were normalised to unity
at the (photospheric)
wavelength point 3925 \AA~and 3942.32 \AA\ in order to give an equal relative scale to
the \ion{Ca}{II} K line cores of different observing dates. 

\begin{figure}
\centering
\includegraphics[scale=0.55]{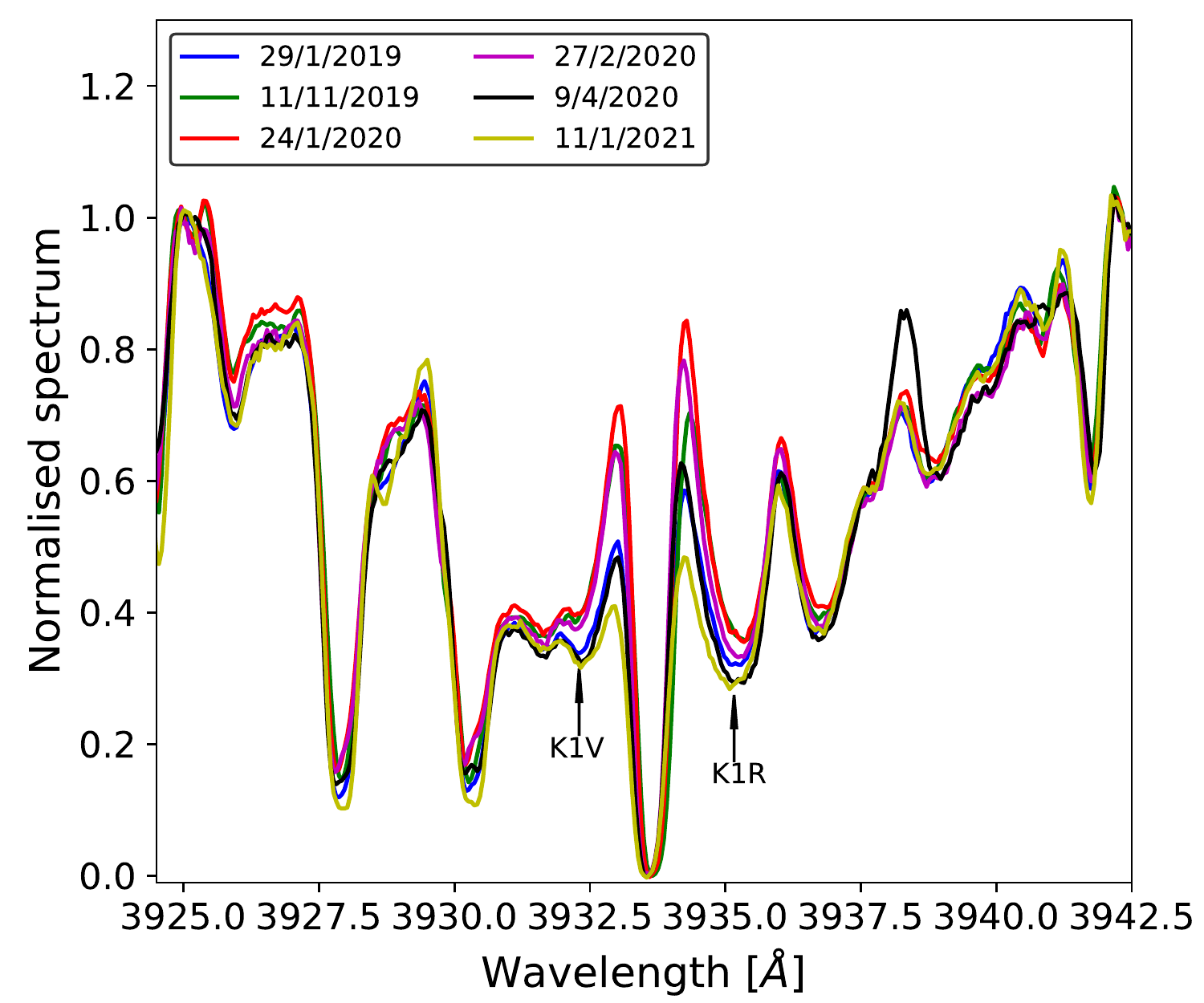}
\caption{Normalised \ion{Ca}{II} K line spectra taken on different dates: 
        29 January 2019 (blue), 11 November 2019 (green), 24 January 2020 (red), 
        27 February 2020 (magenta), 9 April 2020 (black), and 11 January 2021 (yellow).}
\label{CaIIKline}
\end{figure}

In Fig.~\ref{CaIIKline}, we compare these normalised \ion{Ca}{II} K line spectra taken on
29 January 2019 (TIGRE S$_{\rm{II\_MWO}}$=0.554), 11 November 2019 (TIGRE S$_{\rm{II\_MWO}}$=0.697), 24
January 2020 (TIGRE S$_{\rm{II\_MWO}}$=0.800),
27 February 2020 (TIGRE S$_{\rm{II\_MWO}}$=0.748), and 9 April 2020 (TIGRE S$_{\rm{II\_MWO}}$=0.601).
In addition, the TIGRE spectrum of 11 January 2021 (S$_{\rm{II\_MWO}}$=0.492)
serves as a reference for a \ion{Ca}{II} K line core of low activity. This
comparison, at face value, shows that the normalised \ion{Ca}{II} K line flux 
increased until 24 January$^{}$ 2020 and thereafter decreased again. 
The same behaviour is seen in the measured S-values; the method for 
deriving S-values on the Mount Wilson scale is described
in the following subsection (Sect. \ref{sindex_tigre_sec} \& \ref{sindex_tigre_trans}).

Furthermore, we want to draw attention to the  fact that the spectrum 
around the K1V and K1R points is also variable; these points are marked in Fig \ref{CaIIKline}, and form in the
transition from photosphere to chromosphere (see \citep{Vernazza1981ApJS}). These variations may
indicate a changing photospheric contribution,
which is consistent with the results of our temperature study of $\alpha$ Ori; see Sect. \ref{teff_est}. 

\subsection{S-index calculation for giants}

The Mount Wilson program \citep[described in detail by ][]{Wilson1982ApJ} contains
S-index measurements
of giants. These were obtained with a wider bandpass filter for the line cores than the 
triangular bandpass with a FWHM of 1.09 \AA~(used for main sequence stars) 
in order to accommodate, at least to some extent, the widening of the emission lines in giants 
(caused by the so-called Wilson Bappu effect). 
To compare the TIGRE raw S-index measurements of giants (S$_{\rm{II\_TIGRE}}$) with the 
historical Mount Wilson data, the transformation onto the Mount Wilson scale was 
derived from a large set of observations of different giants with a wide bandpass, 
which were also observed by the Mount Wilson group.

\subsubsection{The Mount Wilson S-index for giants}
\label{def_smwo}

The commonly used Mount Wilson S-index is defined as \citep{Vaughan1978PASP90267V}
\begin{eqnarray}\label{sindex_mwo}  
S_{\rm{MWO}} = \alpha \left( \frac{N_{H}+N_{K}}{N_{R}+N_{V}} \right),
\end{eqnarray}     
where N$_{H}$ and N$_{K}$ are the flux counts
in the line cores of the \ion{Ca}{II} H\&K lines in a triangular bandpass with a FWHM
of 1.09 \AA, and N$_{H}$ and N$_{K}$  are the flux counts of two 20 \AA~wide pseudo-continuum 
bandpasses, centred at 3901.07 \AA~and 4001.07 \AA, which serve as a reference 
for the photospheric spectrum. In this fashion, the S-index value obtained is 
independent of the transparency of the night sky,  and the factor $\alpha$ is merely a scaling
factor \citep{Vaughan1978PASP90267V, duncan1991} used to compare different generations of
master-built four-channel S-measurement units on Mount Wilson. 

However, as already mentioned above, the \ion{Ca}{II} H\&K line cores of giant stars are 
wider than those of main sequence (MS) stars. Therefore, \citet{Wilson1982ApJ}
introduced a wider line bandpass for S-measurements of giants.
This band pass has a trapezoidal shape with a FWHM of 1.5 \AA~ and
the top has a width of 1 \AA. However, the wide continua bandpasses 
are the same as those used for the normal S-index \citep{choi1995PASP}. 
For our giant S-value measures on TIGRE spectra, we adopted a rectangular \mbox{2 \AA}~line profile.
This choice loses even less emission from supergiants than the Mount Wilson
trapezoidal profile, and the calibrated transformation handles the different throughput. 

\begin{figure}
\centering
\includegraphics[scale=0.58]{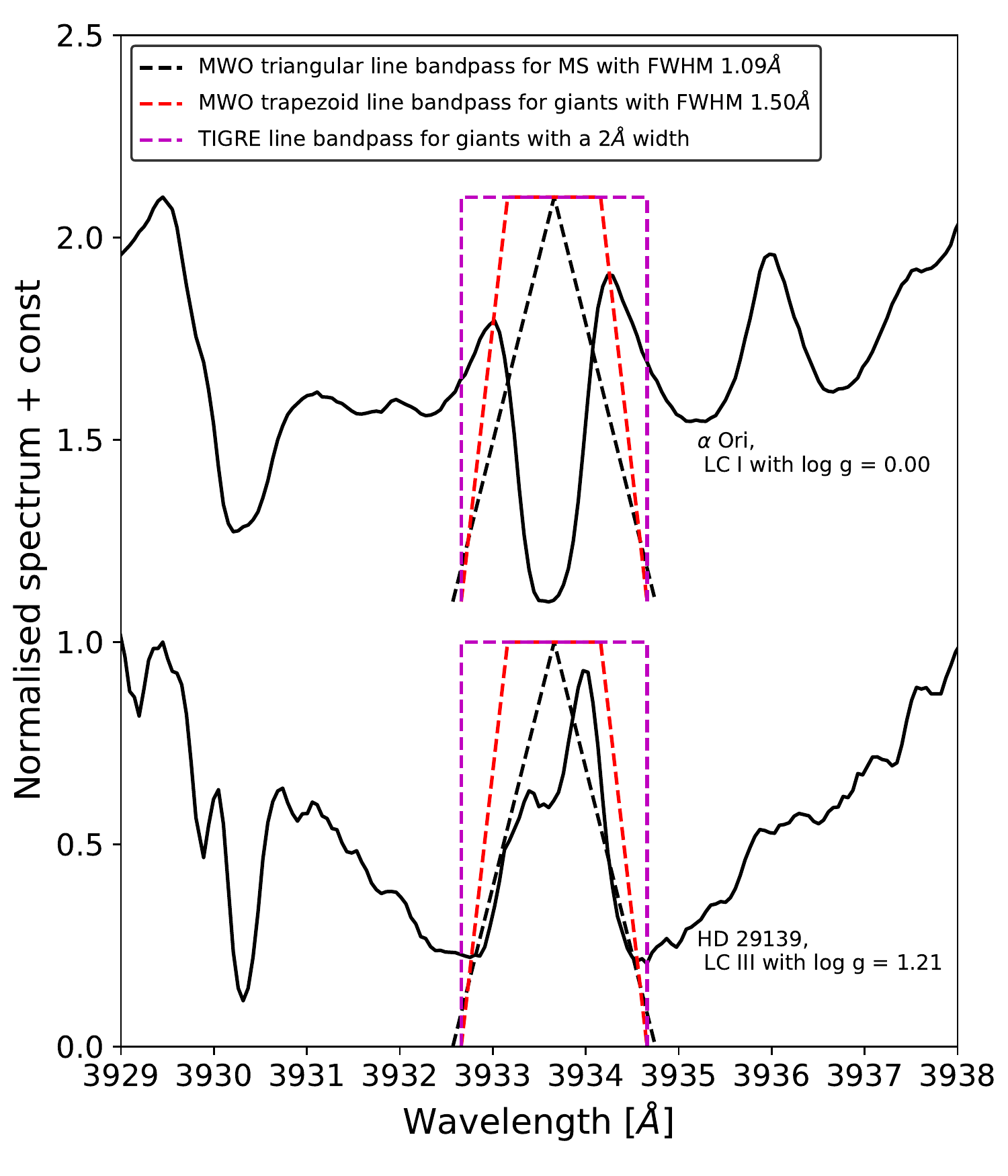}
\caption{TIGRE spectra of the \ion{Ca}{II} K line of two giants (HD~29139 (LC III) 
and $\alpha$ Ori (LC I)) are shown with the line bandpasses of the different 
S-indices: The black dashed line outlines the triangular bandpass, sufficient 
in width for MS stars, and the red dashed line shows the wider
trapezoidal bandpass used for giants. The magenta dashed line shows the rectangular
bandpass used for the raw S$_{TIGRE}$ index; see Sect. \ref{sindex_tigre_sec}.}
\label{LineBandpasses}
\end{figure}

Figure \ref{LineBandpasses} shows the two different
line-core bandpasses of the Mount Wilson S-indices plotted over a TIGRE spectrum 
of the \ion{Ca}{II} K line of the giant HD~29139 ($\alpha$ Tau). The triangular bandpass 
is outlined by the black dashed line, and the wider trapezoidal bandpass by the red 
dashed line.

A  visual comparison of both bandpasses shows that even the wider triangular
bandpass is too small to capture the full line-core emission.   
\citet{Rutten1984A&A} provides a more detailed comparison and discussion
of the S-values obtained with both of these bandpasses; his main conclusion is that 
the wider band pass for index S$_{\rm{II\_MWO}}$ becomes too narrow already for 
luminosity class (LC) I-II giants. The S-index can then only be used for variability monitoring, and is not useful for
quantitative studies. We emphasise this point by superimposing the trapezoidal bandpass
over the \ion{Ca}{II} K line of $\alpha$ Ori, an LC I star, in Fig. \ref{LineBandpasses}.

\subsubsection{The TIGRE S$_{\rm{II}}$-index for giants}
\label{sindex_tigre_sec}

We define the raw, uncalibrated TIGRE S-index for giants in the form
\begin{eqnarray}\label{sindex_tigre}  
  S_{\rm{II\_TIGRE}} = \left( \frac{N_{H}+N_{K}}{N_{R}+N_{V}} \right),
\end{eqnarray}
where $N_{*}$ are the counts in the individual bandpasses.
While we used the same continuum bandpasses ($N_{R}~\rm{and}~N_{V}$) as for the Mount 
Wilson S-index, our TIGRE \ion{Ca}{II} H\&K line bandpasses (for $N_{H}~\rm{and}~N_{K}$) 
are simply rectangular with a width of 2~\AA\  (see Fig. \ref{LineBandpasses}). This choice gives 
the same width at the foot of the profile, but produces less count losses compared to 
the trapezoidal Mount Wilson profile.

\begin{figure}
\centering
\includegraphics[scale=0.4]{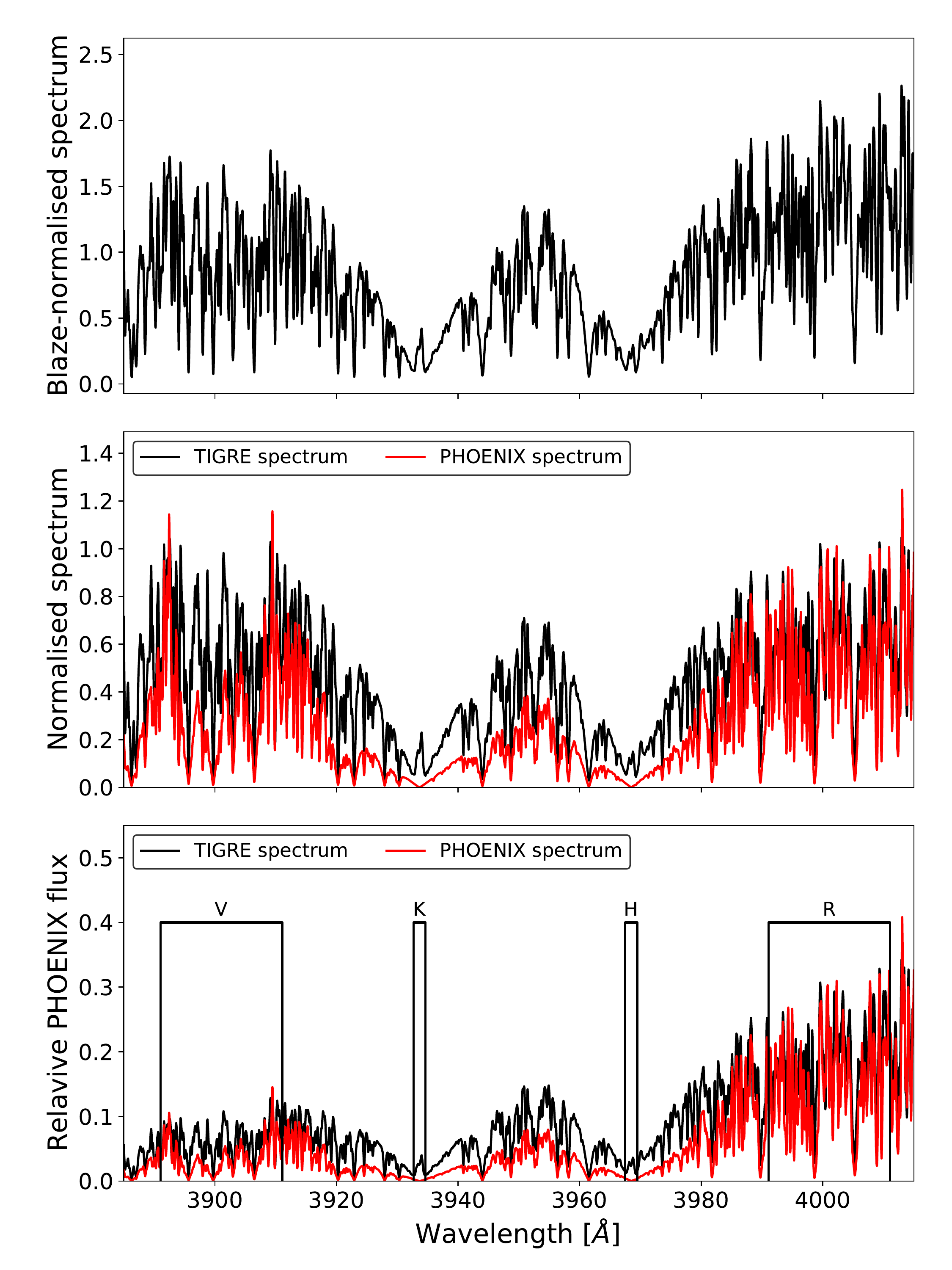}
\caption{Steps of the re-normalisation of spectra for TIGRE S$_{\rm{II\_MWO}}$ estimation: Upper
  panel shows the \ion{Ca}{II} H\&K region of HD~29139 taken with TIGRE on December 27 2019; the middle panel shows the
  continuum-normalised spectra of the TIGRE spectrum as a black line and the selected and continuum-normalised PHOENIX spectrum
  as a red line, and the lower panel shows the re-normalised TIGRE spectrum as a black line
  and the PHOENIX spectrum in a relative flux scale (flux divided by the maximal flux).}
\label{s_cal_steps}
\end{figure}

Measuring the S$_{\rm{II\_TIGRE}}$ index requires three main steps, which are
illustrated in Fig. \ref{s_cal_steps}. The first step concerns the values of the pseudo-continua,
where the calibration fluxes $N(R)$ and $H(V)$ are measured. The standard
normalisation process for TIGRE
spectra may result in incorrect ratios between the fluxes of those reference bands.  
In the upper panel, a normalised TIGRE spectrum of the \ion{Ca}{II} H\&K lines 
is shown. A matching (by effective temperature and gravity) PHOENIX spectrum
of models by \citep{hauschildt1999} is now required to estimate the true flux ratios
between the fluxes of those reference bands. 

We normally use averages of the T$_{eff}$ and log g values in the PASTEL 
catalogue (Version 2020-01-30; \citep{Soubiran2016}), except in the case of the 
variable effective temperature of $\alpha$ Ori. Instead, we adopt the values derived
by our approach. According to those T$_{eff}$ and log~g values, 
the corresponding PHOENIX spectrum is selected from the spectral database \citep{Husser2013A&A}.
To match the observed TIGRE/HEROS spectrum, we consider the rotational line-broadening
(v$\cdot$sin(i)) and a spectral 
resolving power of R$\approx$20000 in the PHOENIX spectrum. The PHOENIX spectrum adapted in this way is then used 
to estimate the real spectral flux slope in the \ion{Ca}{II} H\&K region in the TIGRE spectrum;
this approach was taken from \citet{Hall1995ApJ} and guarantees consistency in 
the flux counts of the large  reference bandpasses of the pseudo-continua, which form 
the denominator of the S-index.

The second step in deriving S$_{\rm{II\_TIGRE}}$ is the assessment of the exact RV
shift of the \ion{Ca}{II} H\&K region of the respective spectrum by means of a cross correlation 
with the PHOENIX spectrum (see middle panel Fig. \ref{s_cal_steps}). Subsequently, the 
TIGRE spectral fluxes are also re-normalised to the physical spectral slope seen 
in the PHOENIX spectrum (lower panel Fig. \ref{s_cal_steps}).
Fortunately, by the wise foresight of O.C. Wilson, the bandpasses of the S-index are symmetric, meaning that any
remaining mismatches of the flux slope and the effective 
temperature are cancelled out in the resulting S-index.   

Finally, the counts in the four bandpasses, depicted in the lower panel 
of Fig. \ref{s_cal_steps}, are obtained by integration,
and the ratio between the sum of the two line bandpasses and the sum of the two continuum
bandpasses is computed according to Eq. \ref{sindex_tigre}. The error 
estimation of S$_{\rm{II\_MWO}}$ follows the method used for S$_{\rm{TIGRE}}$ and 
described by \citet{Mittag2016A&A}. Indeed, apart from the continuum flux 
re-normalisation and line-core profiles, both indices are obtained in a very similar 
manner.

\subsubsection{Transformation of S$_{\rm{II\_TIGRE}}$ to the Mount Wilson scale}
\label{sindex_tigre_trans}

The transformation between the TIGRE and the Mount Wilson giant S-index 
scales takes care of any instrumental effects. However, because of the large epoch difference, a larger
number of giants needs to be considered to average out 
any individual changes in their activity levels. To do so, we chose giants both 
observed by TIGRE and 
contained in the Mount Wilson sample published by \citet{Radick_MWO}.
In total, we use 25 objects to derive the transformation, of which the 
S$_{\rm{II\_MWO}}$ values were measured by the present authors more than ten times.

\begin{figure}
\centering
\includegraphics[scale=0.45]{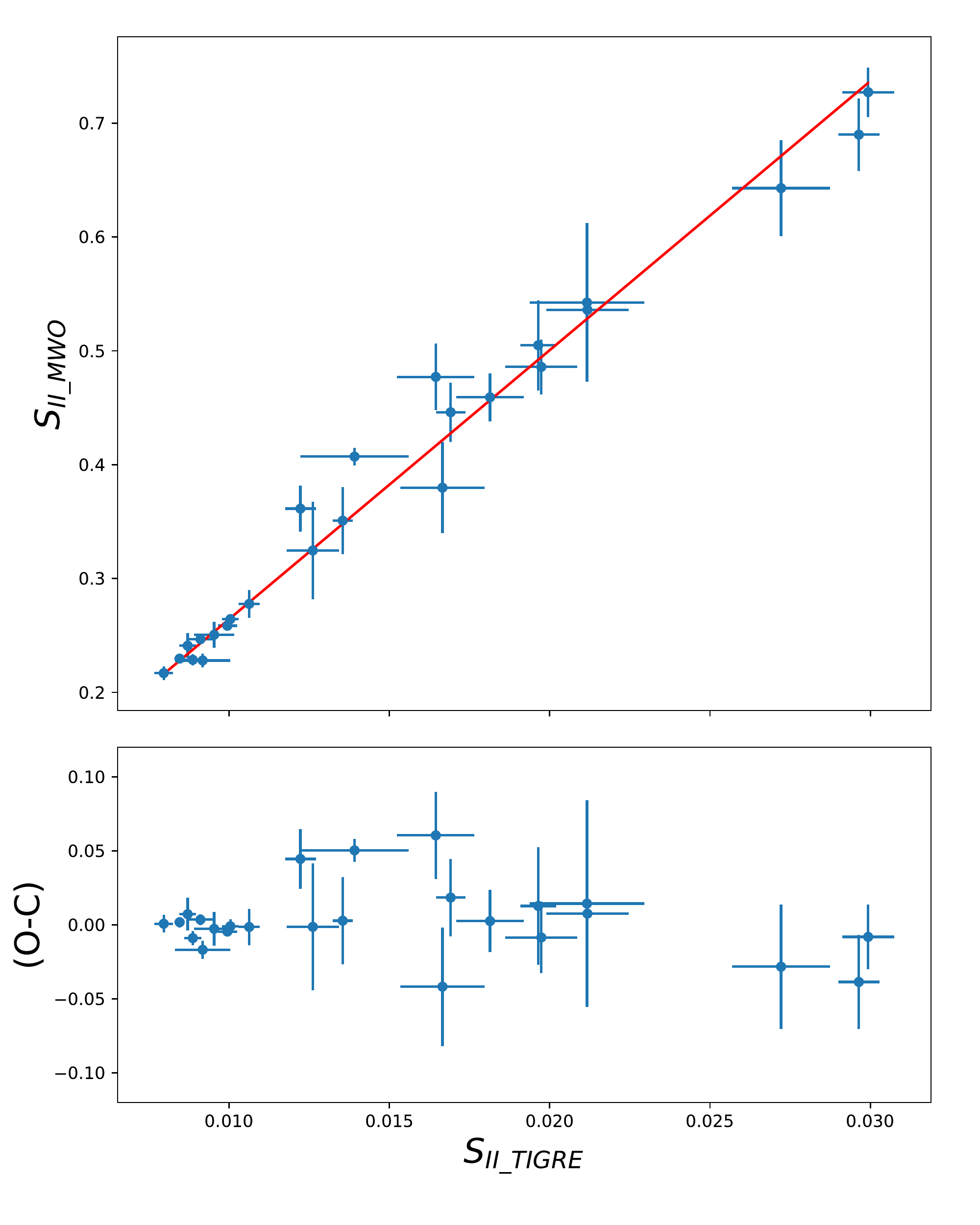}
\caption{S$_{\rm{II\_MWO}}$ vs. S$_{\rm{TIGRE}}$ transformation: Upper panel: S$_{\rm{II\_MWO}}$ vs. S$_{\rm{II\_TIGRE}}$. The solid
  line presents the transformation relation. Lower panel: Residuals of the used transformation relation.}
\label{s_index_transformation}
\end{figure}
\begin{table*}[!t]    
\caption{Giants used for the transformation of the S$_{\rm{II\_TIGRE}}$ 
into the Mount Wilson scale.}    
\label{cal_obj}    
\begin{center}    
\begin{small}
\setlength{\tabcolsep}{5pt}
\begin{tabular}{l c c c c c c c c}
\hline    
\hline
Object & T$_{\rm eff}$ [K]  & $\log$(g)  & S$_{\rm{II\_MWO}}$ & No. of S$_{\rm{II\_MWO}}$ & S$_{\rm{II\_TIGRE}}$ & TIGRE S$_{\rm{II\_MWO}}$ & No. of S$_{\rm{II\_TIGRE}}$ & Difference [$\%$] \\
\hline
\noalign{\smallskip}
HD4128 & 4800 & 2.5 & 0.38$\pm$0.04 & 486 & 0.0167$\pm$0.0013 & 0.422$\pm$0.033 & 29 & 11.0 \\ 
HD6805 & 4600 & 2.5 & 0.229$\pm$0.005 & 35 & 0.0089$\pm$0.0003 & 0.238$\pm$0.011 & 37 & 3.9 \\ 
HD6860 & 3800 & 1.5 & 0.536$\pm$0.01 & 45 & 0.0212$\pm$0.0013 & 0.528$\pm$0.034 & 35 & -1.4 \\ 
HD8512 & 4700 & 2.5 & 0.217$\pm$0.006 & 45 & 0.008$\pm$0.0003 & 0.216$\pm$0.01 & 32 & -0.4 \\ 
HD27371 & 5000 & 2.5 & 0.325$\pm$0.043 & 774 & 0.0126$\pm$0.0008 & 0.326$\pm$0.022 & 57 & 0.4 \\ 
HD27697 & 5000 & 2.5 & 0.251$\pm$0.011 & 777 & 0.0095$\pm$0.0006 & 0.253$\pm$0.017 & 56 & 1.1 \\ 
HD28305 & 4900 & 2.5 & 0.228$\pm$0.006 & 741 & 0.0092$\pm$0.0009 & 0.245$\pm$0.022 & 32 & 7.4 \\ 
HD28307 & 5000 & 3.0 & 0.361$\pm$0.02 & 725 & 0.0122$\pm$0.0005 & 0.317$\pm$0.015 & 33 & -12.3 \\ 
HD29139 & 3900 & 1.0 & 0.446$\pm$0.026 & 204 & 0.0169$\pm$0.0005 & 0.428$\pm$0.016 & 35 & -4.1 \\ 
HD31398 & 4200 & 1.5 & 0.459$\pm$0.021 & 320 & 0.0181$\pm$0.0011 & 0.457$\pm$0.028 & 28 & -0.6 \\ 
HD39801 & 3600 & 0.0 & 0.542$\pm$0.07 & 292 & 0.0212$\pm$0.0018 & 0.528$\pm$0.045 & 73 & -2.6 \\ 
HD71369 & 5200 & 2.5 & 0.258$\pm$0.003 & 398 & 0.0099$\pm$0.0003 & 0.263$\pm$0.011 & 71 & 1.8 \\ 
HD81797 & 4100 & 1.5 & 0.351$\pm$0.029 & 66 & 0.0135$\pm$0.0003 & 0.348$\pm$0.013 & 41 & -0.8 \\ 
HD82210 & 5300 & 3.5 & 0.643$\pm$0.042 & 503 & 0.0272$\pm$0.0015 & 0.671$\pm$0.04 & 79 & 4.4 \\ 
HD96833 & 4600 & 2.5 & 0.23$\pm$0.003 & 44 & 0.0085$\pm$0.0002 & 0.228$\pm$0.009 & 69 & -0.8 \\ 
HD109379 & 5100 & 2.5 & 0.264$\pm$0.005 & 99 & 0.01$\pm$0.0003 & 0.265$\pm$0.011 & 42 & 0.4 \\ 
HD111812 & 5600 & 3.0 & 0.727$\pm$0.022 & 322 & 0.0299$\pm$0.0008 & 0.735$\pm$0.027 & 32 & 1.1 \\ 
HD115659 & 5100 & 2.5 & 0.278$\pm$0.012 & 331 & 0.0106$\pm$0.0003 & 0.279$\pm$0.012 & 67 & 0.5 \\ 
HD164058 & 3900 & 1.5 & 0.407$\pm$0.008 & 15 & 0.0139$\pm$0.0017 & 0.357$\pm$0.041 & 23 & -12.4 \\ 
HD124897 & 4300 & 1.5 & 0.247$\pm$0.004 & 917 & 0.0091$\pm$0.0004 & 0.243$\pm$0.012 & 72 & -1.4 \\ 
HD159181 & 5200 & 1.5 & 0.69$\pm$0.032 & 611 & 0.0296$\pm$0.0006 & 0.728$\pm$0.024 & 70 & 5.6 \\ 
HD186791 & 4200 & 1.5 & 0.505$\pm$0.04 & 461 & 0.0196$\pm$0.0006 & 0.492$\pm$0.019 & 28 & -2.5 \\ 
HD202109 & 4900 & 2.5 & 0.241$\pm$0.011 & 222 & 0.0087$\pm$0.0003 & 0.234$\pm$0.01 & 68 & -3.0 \\ 
HD205435 & 5100 & 3.0 & 0.477$\pm$0.029 & 461 & 0.0164$\pm$0.0012 & 0.417$\pm$0.031 & 66 & -12.7 \\ 
HD209750 & 5200 & 1.5 & 0.486$\pm$0.024 & 543 & 0.0197$\pm$0.0011 & 0.494$\pm$0.03 & 52 & 1.7 \\ 
\hline
\end{tabular}
\tablefoot{The listed T$_{\rm eff}$ [K] and $\log$(g) are the values for the PHOENIX spectrum that used for the TIGRE
  S-index estimation.} 
\end{small}
\end{center}
\end{table*}
These giants ---with their physical parameters and median S values, with the standard deviation of the median as error---
are listed in Table \ref{cal_obj}. The rounded T$_{\rm eff}$
and log~g values are those of the selected PHOENIX reference spectra. 

Figure \ref{s_index_transformation} gives the distribution of  S$_{\rm{II\_MWO}}$ vs. 
S$_{\rm{II\_TIGRE}}$ (average values)  for all 25 giants.
Here, a linear trend is very obvious and we derive the following
transformation relation via an orthogonal distance regression, which is represented in
the same figure by a solid line:
\begin{eqnarray}\label{transformation_eq}  
  S_{\rm{II\_MWO}} = (0.0280\pm0.0063)+(23.63\pm0.61)\rm{S}_{\rm{II\_TIGRE}},
\end{eqnarray}
which yields a standard deviation of the residuals of 0.024. 
In addition, we compute the percentage deviations (listed in Table \ref{cal_obj})
between the Mount Wilson reference S-values and the transformed TIGRE S-values, 
and obtain an averaged deviation of 5.6\%. This is mainly
an effect of long-term changes in the activity levels between the very different 
epochs of the observations, but is also attributable, to a much lesser extent, to observational error.

\subsection{Long-term time series: joining TIGRE and Mount Wilson S-index values}
\label{S-index_time_series}

We now exploit the fact that the red supergiant $\alpha$ Ori is both part of our 
TIGRE giants observation program and contained in the Mount Wilson 
program (see \citet{Radick_MWO}), with available data reaching back into the 1980s.
These data allow us to examine whether or not a similar behaviour of the S-index as observed during 
the great dimming in the winter between 2019 and 2020 ---with a simultaneous significant increase in the S-index
anti-correlated with the V-band luminosity--- has already occurred before.

\begin{figure*}
\centering
\includegraphics[scale=0.6]{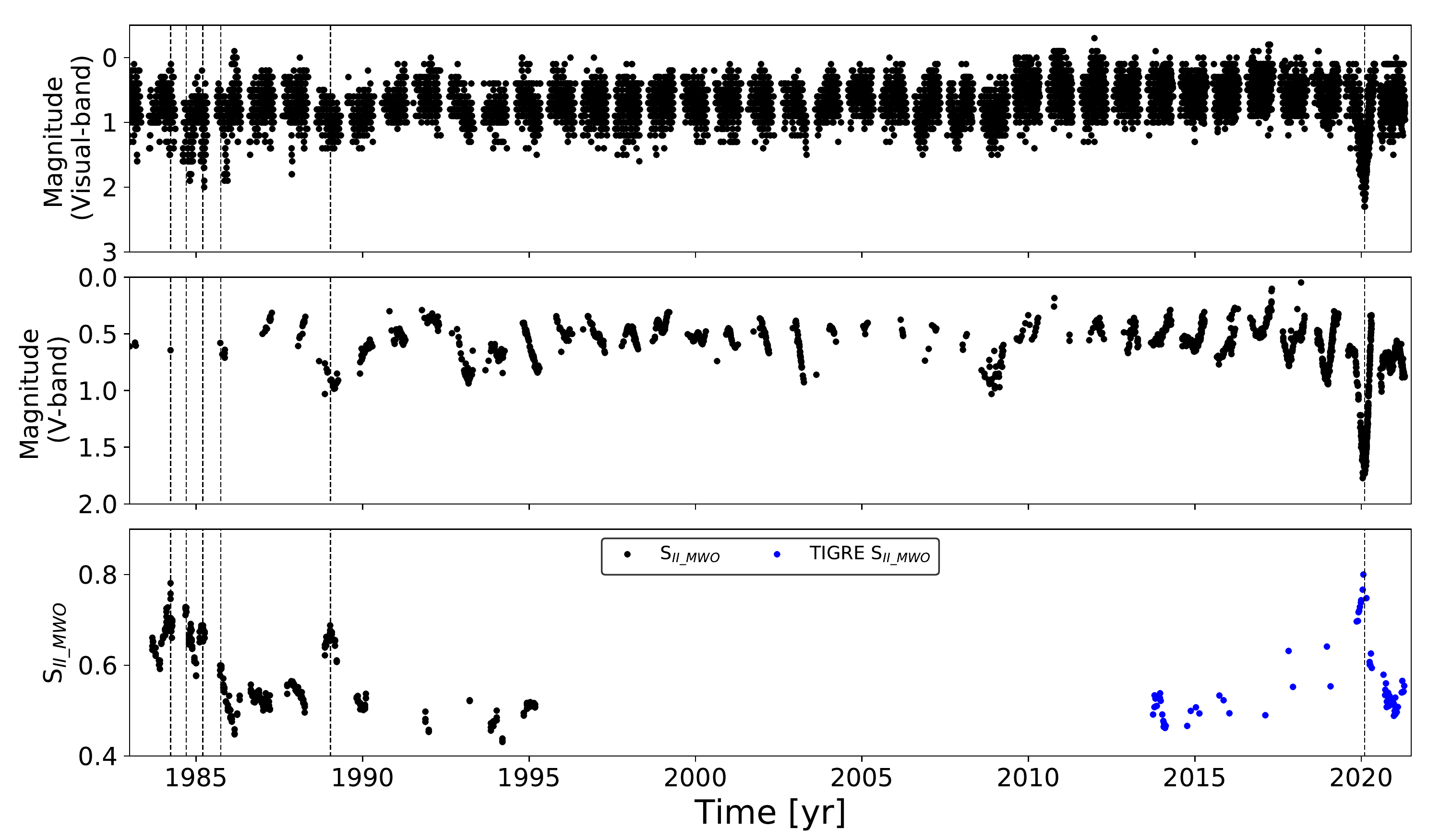}
\caption{Long-term time series: Comparison of the visual (upper) and V band (middle)
magnitudes of the AAVSO database with S-values of the Mount Wilson scale (lower panel). 
The dashed lines mark the periods with increased S-values.}
\label{sindex_aavso}
\end{figure*}

The combined, long-term S-index time series is shown in the lower panel of 
Fig. \ref{sindex_aavso}.   The Mount Wilson data are represented by black points 
and cover the period of September 1983 to March 1995. The TIGRE observation of $\alpha$ Ori 
started in September 2013 once TIGRE went into operation.
Only S-index values based on spectra with an S/N larger
than 25 at 4000 \AA~are shown (as blue points). In total, a period of 38 years 
is covered, from 1983 to 2021, albeit with a gap of 
18 years between 1995 and 2013. As can be seen from Fig. \ref{sindex_aavso},
the median of the $\alpha$ Ori S-value is 0.539 over the whole period.

At least three episodes of strongly increased S-index values occurred during 
those 38 years. The first one took place around 1984 with a maximum S-value of 0.7811 
on 25 March, 1984, which is  45\% larger than the long-term median S-value,
with a phase of increased S-index level continuing until 1985. The second event took place in 
1989 with a maximum S-value of 0.6879 on 1 October, 1989, which is 28\% above the median 
S-index value. The third and most significant event is the `great dimming'  in 2020
with a maximum 
S-index value of 0.800 reached on 24 January, 2020, which is  48\% larger than
the median S-value of 0.539. To see whether the earlier two events are of a 
similar nature to the 2020 event, we now need to look at the correlation with photometric data.

\subsection{Comparing S-index and photometric variations}

The latest 2020 increase in S-index values showed a good anti-correlation and 
synchronisation with the great dimming event in 2020. To take a look further back 
in time, we again employ the photometric data taken from the AAVSO database.
Figure~\ref{sindex_aavso} shows the light curve of the visual band magnitude in the upper
panel, and the V band magnitude (Johnson V-band) versus time in the middle panel.
To ease this comparison, the above-mentioned periods of increased S-values are
marked by dashed vertical lines. The visual magnitude light curve does
show a dimming
in 1985, slightly delayed against the maximum increase in S-index values around 1984. 
However, we caution that the decrease in the visual magnitude
in 1984 and 1985 was established by only two observers and very few brightness estimates were made.  
However, the anti-correlation with the S-index supports the idea that we  have 
an event similar to the great dimming in the winter between 2019 and 2020.

Unfortunately, V-band photometric data of before $\approx$1988 are quite rare, and so
we need to use the more abundant visual magnitude estimates for our comparison 
with the older S-index time series. In the case of the S-increase around 1 October, 1989, the
decrease in visual brightness is not as strong as during the 2020 great dimming 
event, but neither is the increase in S-index in 1989. From the few V-band data available, 
the brightness decrease can be confirmed and can therefore be considered as real.  
As mentioned above, the evidence for the 1984 and 1985 variation is less clear, yet
there seems to be a recurrent pattern; namely that when the S-index rises, the visual brightness
decreases. Such events last only $o(1)$yr, but lie several or even many years 
apart. 

A simple explanation for this anti-correlation of S-index and V-mag lies in the 
definition of the S-index, according to which the near-ultraviolet (NUV) continuum fluxes of 
the reference windows form the 
denominator. As now seen very clearly in the recent great dimming event, when 
the brightness, and with it the temperature, decrease (see \citet{Gray2008AJ},
\citet{Harper2020ApJ} and this work), the denominator of the S-index formula 
decreases, and consequently the S-index of $\alpha$ Ori increases, even if the chromospheric 
emission remains constant on an absolute scale. In addition, \citet{Dupree2020ApJ}
present radial velocity data that indicate a variation of the stellar radius, adding to the 
effect of reduced continuum flux during such a dimming event.
All these observations show that the photosphere of $\alpha$ Ori is not at all
stable, and that in such a case the S-index is as sensitive to photospheric 
variations as it is to chromospheric emission changes. 

\subsection{The absolute scale: \ion{Ca}{II} H\&K flux}
\label{ca_II_hk_flux_est}

To obtain the true activity level of $\alpha$ Ori based on the S-index, the latter has 
to be transformed into physical emission line fluxes on an absolute scale. 
Given the photospheric variability, there is no single relation
for $\alpha$ Ori to achieve this. Instead, we need to consider all components of S, the 
absolute NUV-continuum fluxes, and the line emission individually and for each spectrum separately.

To accomplish this we use the method developed by  \citet{Linsky1979ApJS}, which was originally
designed to derive the absolute \ion{Ca}{II} H\&K flux for $\alpha$ Ori.
However, this approach requires the removal of the instrumental and atmospheric
effects on the spectral flux distribution. For this purpose, part of the TIGRE 
observing schedule on each night is an observation of at least one spectrophotometric
standard star to create the instrumental (and atmospheric) response function for 
each night. Hence, for 73 $\alpha$ Ori spectra, we were actually able to obtain this
spectrophotometric calibration.

\begin{figure}
\centering
\includegraphics[scale=0.4]{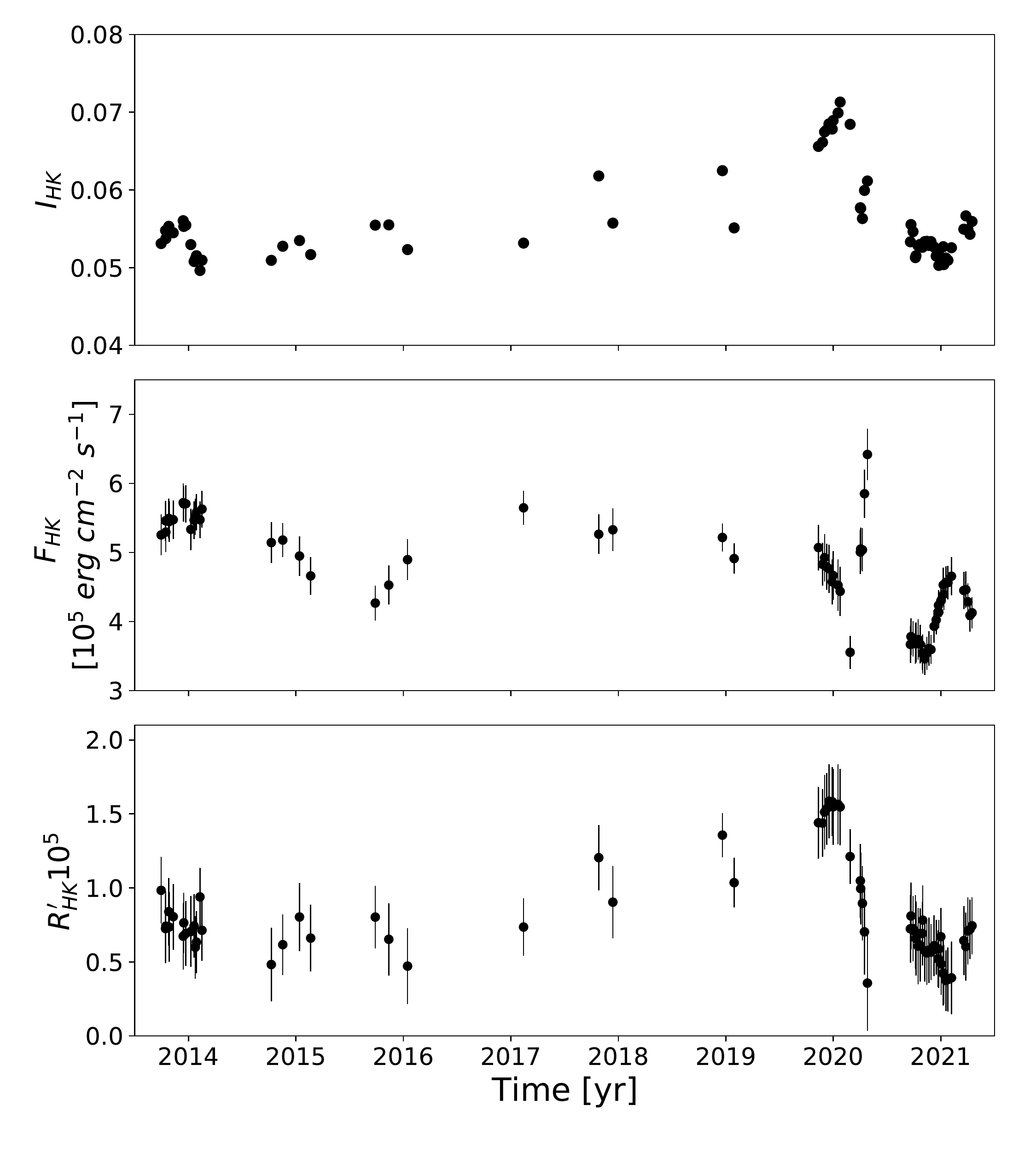}
\caption{\ion{Ca}{II} H\&K fluxes in relative and absolute flux scale: Upper
  panel shows the count ratio $I_{HK}$ vs. time, middle panel the
  estimated absolute \ion{Ca}{II} H\&K flux, and the lower panel the
  estimated \ion{Ca}{II} H\&K flux excess $R^{\prime}_{HK}$.}
\label{ca_II_hk_flux}
\end{figure}

After applying the nightly instrumental response function, the count ratio is
calculated for observations with a sufficient S/N (larger 
than 25 at 4000 \AA), following the definition by \citep{Linsky1979ApJS}, that is,
\begin{eqnarray}\label{linsky_flux_count_ratio}  
 I_{HK}  = \left( \frac{N_{H1}+N_{K1}}{N_{3925-3975}} \right),
\end{eqnarray}
where $N_{H1}$ and $N_{K1}$ are the spectral flux counts between the so-called K1 and H1 points 
of the \ion{Ca}{II} H\&K lines, respectively, and $N_{3925-3975}$ is the count in a
continuum bandpass of  50~\AA~in width centred at 3950 \AA. In the case of $\alpha$ Ori,
we assumed a distance of 3 \AA~between the K1 and H1 points. The positions of the K1 points
of the \ion{Ca}{II} K line are labelled in Fig. \ref{CaIIKline}. The H1 point positions of the
\ion{Ca}{II} H line are similar to those for \ion{Ca}{II}. In total, we were able to 
derive this count ratio $I_{HK}$ for 72 TIGRE spectra, which are shown in the upper 
panel of Fig. \ref{ca_II_hk_flux}.   

Not surprisingly, given the similarities in definition, the $I_{HK}$ time line shows a
similar behaviour as the S-index values in Fig. \ref{sindex_aavso}. After all, both indices 
are based on a ratio between the flux counts in the H\&K line cores and a UV
reference continuum window, with the latter amplified by a change of the stellar 
radius, changing with a high power with decreasing effective temperature 
during the great dimming event.

To convert the $I_{HK}$ values into the absolute \ion{Ca}{II} H\&K fluxes, we 
then follow the relation given by \citep{Linsky1979ApJS}, that is, 
\begin{eqnarray}\label{linsky_flux}  
\mathcal{F}_{\rm{H\&K}} =   I_{HK} \mathcal{F}(\Delta \lambda)*50,
\end{eqnarray}
where, $\mathcal{F}(\Delta\lambda)$ relation is given by \citet{Linsky1979ApJS}:
\begin{eqnarray}\label{linsky_const_flux}  
\mathcal{F}(\Delta\lambda) = 8.264-3.076(V-R)
.\end{eqnarray}
To do so, we perform a transformation of the effective temperatures (as derived above) 
into $V-R$ colours using Eq. \ref{teff_con_vr}; see Sect. \ref{ca_II_flux_est_discribtion}.
The result is shown in the middle panel of Fig. \ref{ca_II_hk_flux}.   
Comparing these two panels, one immediately notices the entirely
different behaviours of the two time series: while $I_{HK}$-values (just like the S-indices) increase during
the great dimming event, the derived absolute fluxes decrease, and their temporal behaviour 
is more comparable with that of the effective temperature; see Appendix \ref{var_stellar_radius}. 

In this context, we must recall that some part of the H\&K line-core emission is the
underlying photospheric contribution, which ---like the UV reference fluxes--- 
strongly decreased during the great dimming event as well, driven by the decreasing
effective temperature and the stellar radius. To evaluate the pure chromospheric 
activity of $\alpha$ Ori on an absolute scale, the latter has to be separated from 
the photospheric contribution and to be normalised by the corresponding bolometric 
flux. 

This type of activity index was introduced by \citet{Linsky1979ApJS}, who defined 
an $R^{\prime}_{\rm{HK}}$-index through
\begin{eqnarray}
R^{\prime}_{\rm{HK}} & = &  \frac{\mathcal{F}_{\rm{HK}}-\mathcal{F}_{\rm{HK,phot}}}{\sigma~T_{\rm{eff}}^{4}}.
\end{eqnarray}
A brief description of the estimation of photospheric flux ($\mathcal{F}_{\rm{HK,phot}}$) is
given in Sect. \ref{ca_II_flux_est_discribtion},
and the resulting $R^{\prime}_{\rm{HK}}$ time series is presented in the lower panel
of Fig. \ref{ca_II_hk_flux}. Again, there is a remarkable difference compared to the two panels 
above. From the beginning of the dimming event to its brightness minimum,
$R^{\prime}_{\rm{HK}}$ remains constant within its uncertainties. Then, with the 
recovery of the brightness of $\alpha$ Ori, $R^{\prime}_{\rm{HK}}$ (meaning the pure \ion{Ca}{II} H\&K 
flux, or line core `flux excess') decreases. 
It therefore appears that any variation of the pure chromospheric emission,
and so the magnetic heating, started only after the great dimming event had reached
its full strength;  however, given the
large errors of $R^{\prime}_{HK}$, caused mainly by the uncertainty in the effective
temperatures, this result has to be regarded with some caution. 

\subsection{The space view: \ion{Mg}{II} h\&k flux}

Valuable additional observations of the chromospheric activity during the great 
dimming of $\alpha$ Ori were presented by \citet{Dupree2020ApJ}, who
performed measurements of spatially resolved \ion{Mg}{II} h\&k lines 
and MUV fluxes in 2400-2700 \AA~taken on eight different days in 2019 and 2020 
with the HST/STIS. These data suggest a strong increase 
in flux in the period of 18 September to 28 November 2019, with a maximum on 6 October$^{}$ 2019. 

\begin{figure}
\centering
\includegraphics[scale=0.4]{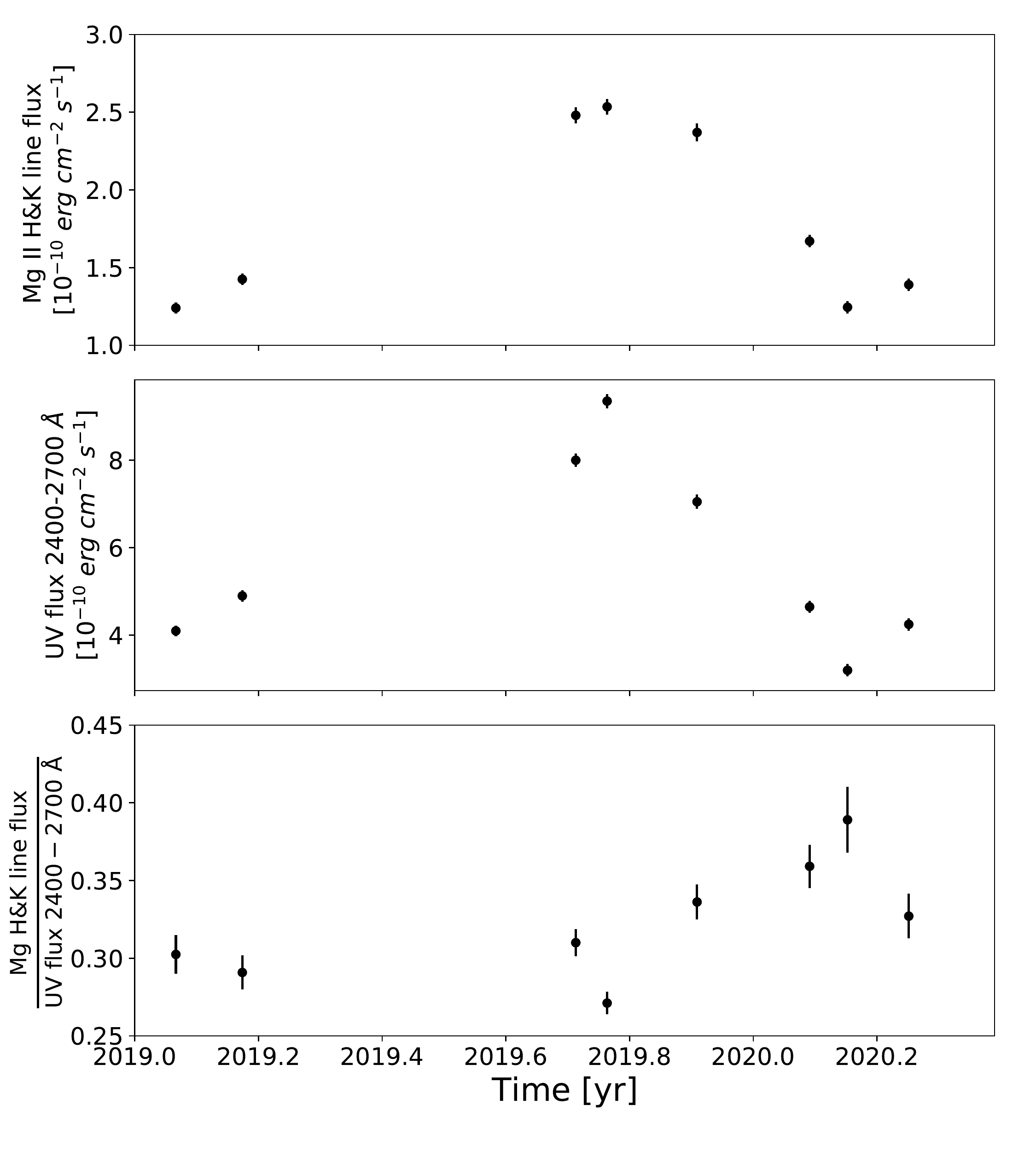}
\caption{Summed \ion{Mg}{II} h\&k spectral line flux (upper panel),
the summed MUV fluxes in 2400-2700~\AA~(middle panel), 
both as obtained from Fig. 5 of \citet{Dupree2020ApJ}, and their ratio (lower
panel).}
\label{mg_II_hk_flux_dupree}
\end{figure}

Unfortunately, individual flux values are not listed in the paper by \citet{Dupree2020ApJ}, but
these values can be reasonably estimated (within their measurement uncertainties)
from Fig. 5 of \citet{Dupree2020ApJ}. The summed \ion{Mg}{II} h\&k line and 
summed MUV fluxes in 2400-2700~\AA~for each day, as well as 
their ratio,  are shown here in the upper, middle, and lower panels of Fig. \ref{mg_II_hk_flux_dupree},
respectively. As expected, the upper panel reproduces the increase in the \ion{Mg}{II} h\&k 
line emission described by \citet{Dupree2020ApJ}; this is similar to the
behaviour of our derived \ion{Ca}{II} H\&K fluxes over time (see middle panel of Fig. \ref{ca_II_hk_flux}). 

Forming the ratio of the \ion{Mg}{II} h\&k line emission over the MUV flux in 2400-2700~\AA~should
then have a similar effect, as seen in the S-index and $I_{HK}$ 
time series, because in principle the photospheric flux part in the measured MUV flux in 2400-2700~\AA, much
like the UV flux in the reference bands of the S-index, is strongly temperature dependent.

Indeed, the lower panel of Fig. \ref{mg_II_hk_flux_dupree} does also show an increase in the \ion{Mg}{II} h\&k
line flux relative to its MUV flux in 2400-2700~\AA~during the great dimming event. 
This demonstrates that there is no discrepancy between the observational evidence presented here regarding 
the chromospheric \ion{Ca}{II} H\&K line emission and the respective \ion{Mg}{II} h\&k
data presented by \citet{Dupree2020ApJ}. Both emissions should be, and indeed are, 
behaving in a very similar way during the great dimming event. The only difference  
is seen in the evidence for changes in the effective temperature,
which is an issue we discuss below.

\section{Dynamical behaviour of the photosphere}

We also detected  significant variations and shape differences 
in other spectral lines:  here we specifically focus on the
wavelength range from 6251~\AA~to 6263.4 \AA, which contains the \ion{V}{I} 6251.82~\AA~and 
\ion{Fe}{I} 6252.56~\AA~lines; the line-depth ratio of these lines depends on 
the effective temperature in cool giants \citep{Gray2001PASP}.

\begin{figure}
\centering
\includegraphics[scale=0.5]{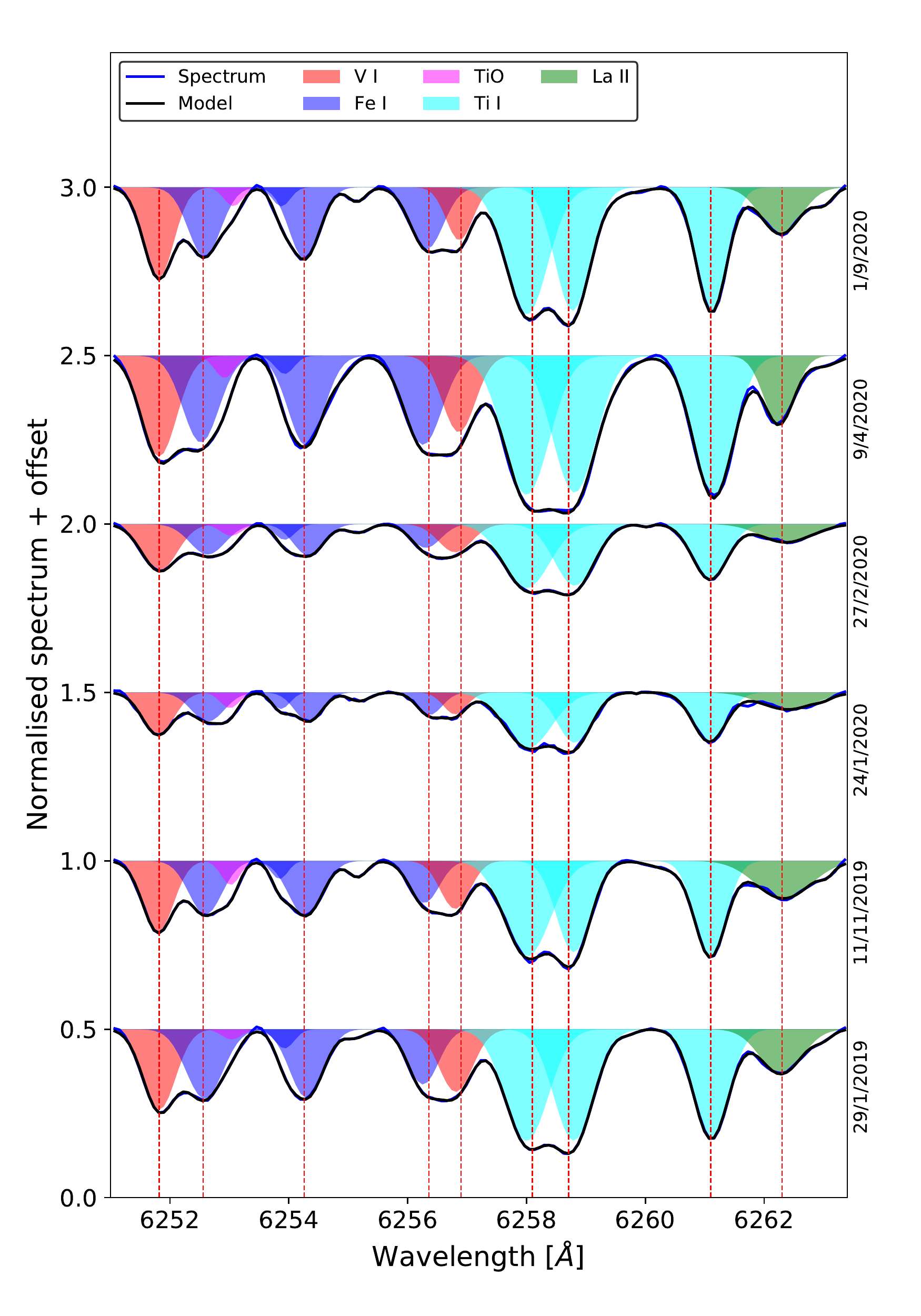}
\caption{Normalised spectra in the wavelength range from 6251 \AA~to 6263.4 \AA~taken before, during, and after
the great dimming event. The different colours mark the different elements and 
the red dashed lines indicate the laboratory wavelengths. }
\label{spec_norm_example}
\end{figure}

First, the spectra were normalised and RV corrected as
described in Sect \ref{RV_var}. In Fig. \ref{spec_norm_example}, we compare the 
normalised spectra  taken before, during, and after the great dimming 
event to demonstrate the spectral changes.
Figure \ref{spec_norm_example}  also demonstrates
that the degree of blending of the involved spectral lines 
varies with time. To disentangle those changes, we performed a multiple Gaussian fit to better
determine the true line centres as well as the width and  
the depth of each line; to ensure robust results we considered only low-noise 
spectra with a mean S/N of at least 100 in the considered wavelength range.
From the derived true line centres, and 
using the laboratory wavelengths as listed in
the NIST Atomic Spectra Database \citep{NIST_ASD}, we then determined the respective
rest frame RVs, which are presented in this section.  

\subsection{Mean RV variations}
\label{RV_var}

First, we had to determine a reference photospheric RV value to which we
compare each $\alpha$ Ori spectrum; for this, we used the line of \ion{Fe}{I} at 6254.26 \AA. 
The RV values for the other spectra were then determined by cross
correlation with that iron line in the first spectrum, and we estimate the uncertainty of this procedure
to be about $\pm$0.4 km s$^{-1}$ based on the assumption that the shift between two
spectra can be determined with a sub-pixel accuracy of $\pm$0.1 pixel. 

\begin{figure}
\centering
\includegraphics[scale=0.75]{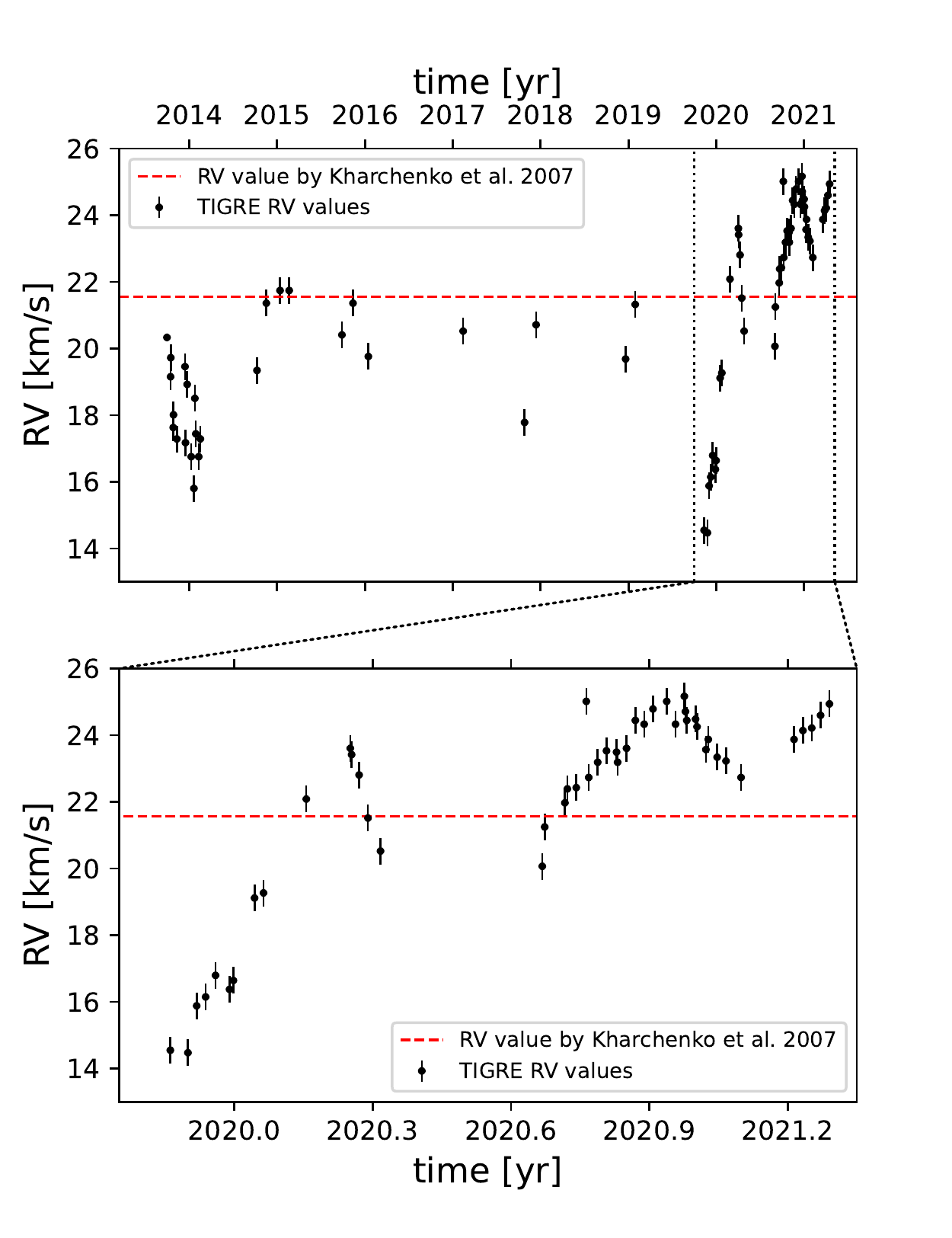}
\caption{Estimated RV values vs. time. The red dashed line marks the long-term average RV
value for $\alpha$ Ori of 21.56$\pm$0.58 km s$^{-1}$ given by \citep{Kharchenko2007AN}.}
\label{rv_time_series}
\end{figure}

In Fig. \ref{rv_time_series},
we plot the resulting RV values.  For the observing season between 2019 and 2020, these RVs are consistent with the
RV values shown by \citet{Dupree2020ApJ}.  Additionally, we plot the average RV value
of $\alpha$ Ori of 21.56$\pm$0.58 km s$^{-1}$ given by \citep{Kharchenko2007AN}, in order to better visualise the RV variations. 
As evident from Fig. \ref{rv_time_series}, 
the RV variations range from $\sim$14 km s$^{-1}$ to $\sim$26 km s$^{-1}$, and we 
note that for the observation season from 2020 to 2021, the RV values are 
larger than the long-term average RV, except for the first two values, 
indicating a redshifted  photosphere until the end of the year 2020. 
Subsequently, the RV values decrease until approximately 2021.1 and we see a blueshifted photosphere;
later, the RV values increase again.

\subsection{Line RV variations}
\label{rest_rv_var}

Visual inspection of the spectra, as shown in Fig. \ref{spec_norm_example} 
reveals some variations of the line centres. Some are blueshifted and others 
are redshifted. These variations include shifts of one line core with respect to another,
suggesting a non-uniform motion of their slightly different geometrical origins. 

In order to gauge the reality of this dynamical picture, we first checked the stability 
of the wavelength solution of our TIGRE spectra in this wavelength range. To do so,
we determined the line centres of various lines in this wavelength 
region with respect to the RV standard star of that night, which is always taken for quality-control
purposes (cf. the TIGRE webpage\footnote{https://hsweb.hs.uni-hamburg.de/projects/TIGRE/DE/\\ \hspace*{1.15cm} hrt\_user/spec\_redu\_info\_rv\_log.php}). 
Each line centre was determined via a multiple Gaussian fit, in the same way as done
for comparing the $\alpha$ Ori lines to one another  as described above.

With respect to the RV standard stars, we find no significant trends in the lines 
cores, confirming the stability of our wavelength solution. The standard
deviation of the rest RV variation of the respective lines (with a
line depth of at least one-tenth of the normalised continuum) suggests a small scatter, 
ranging from 0.07 km s$^{-1}$ for the \ion{Fe}{I} line at 6254.26 \AA~to
0.55 km s$^{-1}$ for the \ion{Ti}{I} line at 6261.10 \AA.
We also checked in this fashion the stability of the line distances; here, we find a 
scatter of less than 0.02 \AA.

\begin{figure}
\centering
\includegraphics[scale=0.57]{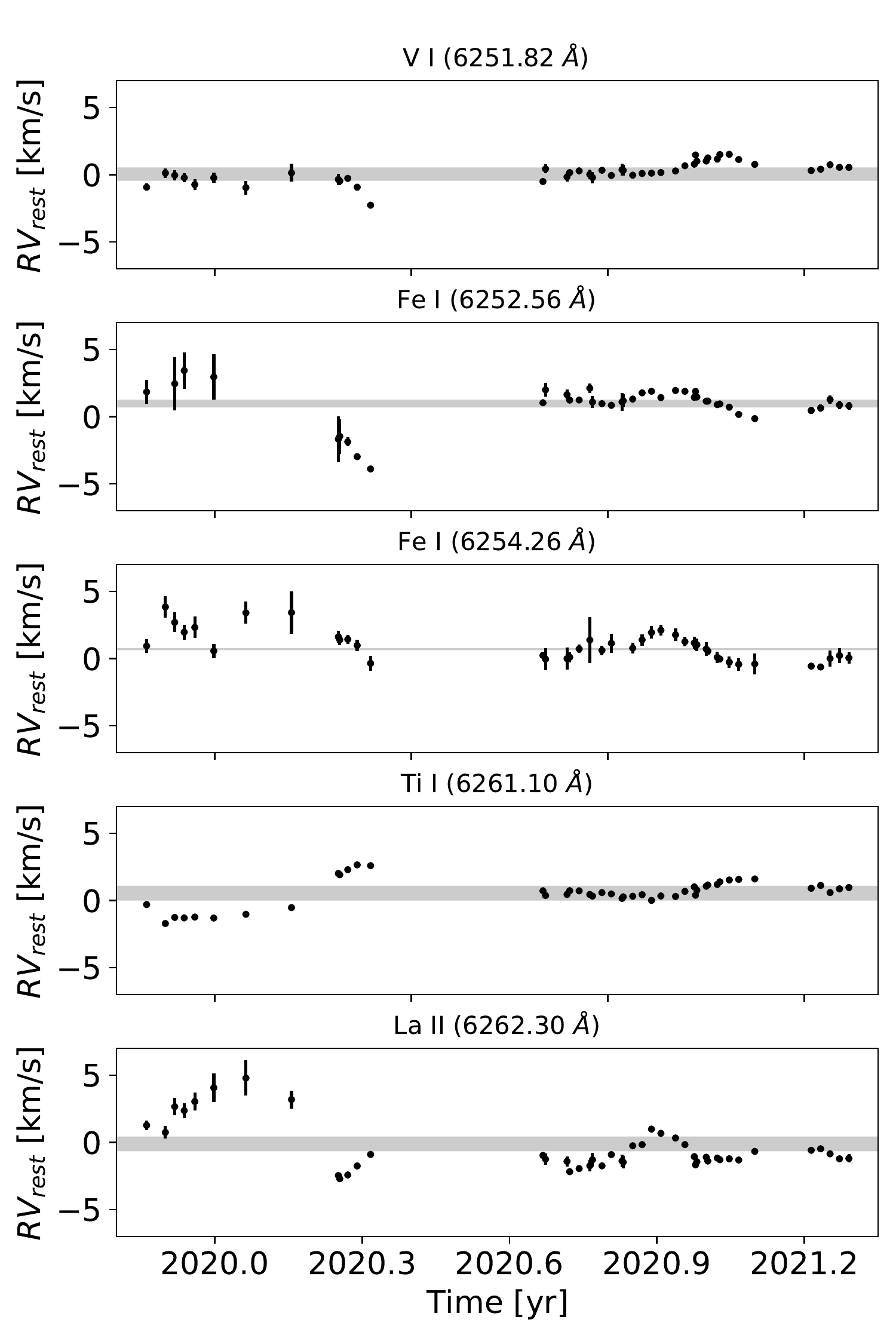}
\caption{Rest RV values over observing epoch. The grey area marks the range of 
the scatter as expected from the line RV examination with respect to 
the RV standard stars.}
\label{RV_res_time_series}
\end{figure}

Following this quality check, we restricted our analysis to spectra with rest-frame
RV errors of below 2 km s$^{-1}$, and considered only RV data more recent than epoch 
2019.5, because the earlier observing cadence was not sufficiently dense. 
The derived time series of rest-frame RV (hereafter, rest RV) values of the \ion{V}{I} at 6251.82~\AA, \ion{Fe}{I} lines
at 6252.56~\AA~and 6254.26~\AA, 
\ion{Ti}{I} line at 6261.10~\AA,~and \ion{La}{II} line at 6262.30~\AA~are shown in Fig. \ref{RV_res_time_series}.
In these time series, we observe some residual variation in the rest RV values.
Also, these time series show a different trend in the rest RV values.
To check whether these variations could be caused by residual errors in the wavelength solution
of the TIGRE \'Echelle spectra, we check the
line distances in this wavelength range in the RV standard stars
observed on the same nights. Here, we do not find any significant scatter in the line distances 
and therefore we conclude that the observed variations are not caused by an erroneous wavelength solution.

\begin{figure}
\centering
\includegraphics[scale=0.45]{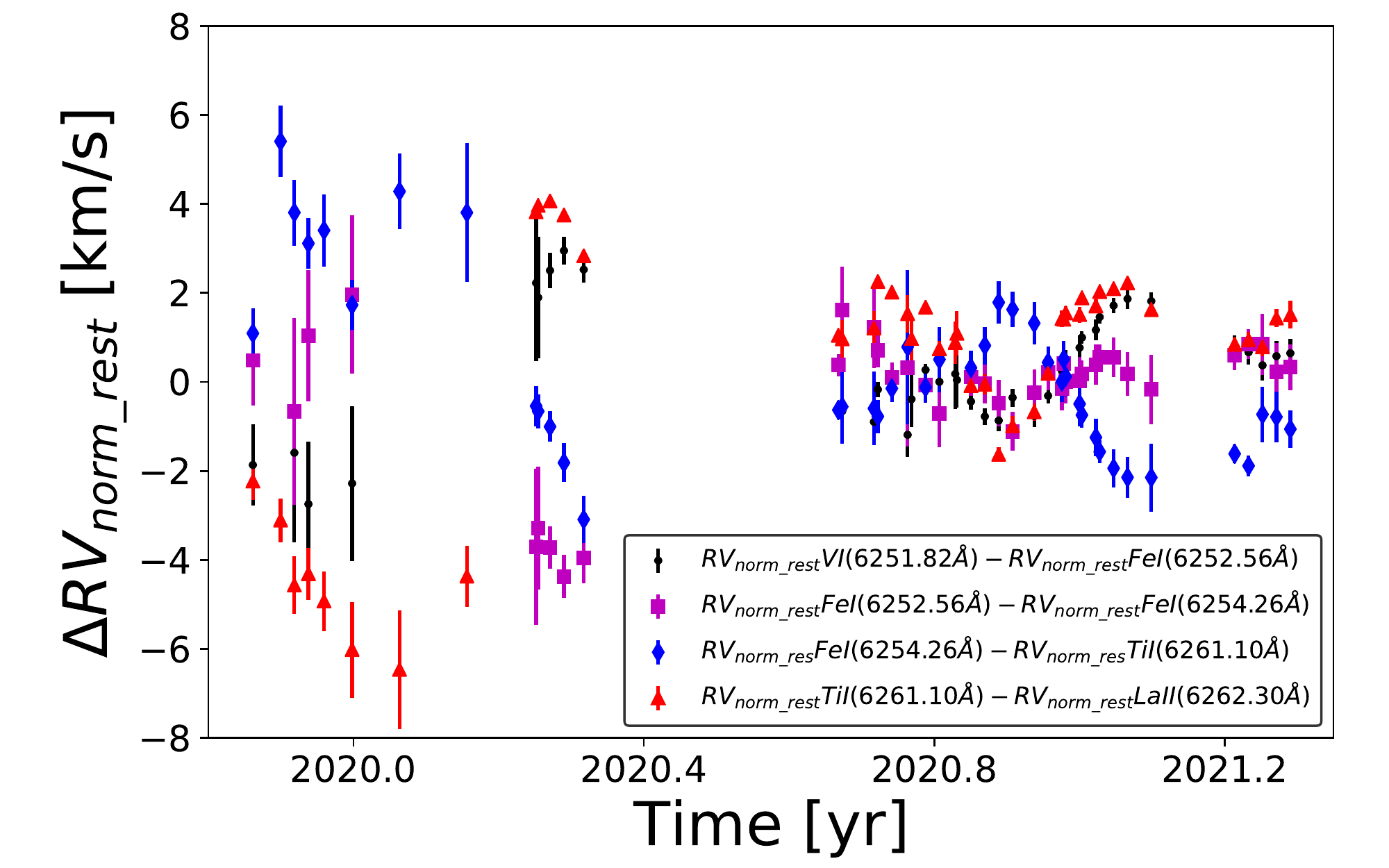}
\caption{Normalised rest RV value vs. time. The different signs and colour codes mark the different line pairs. }
\label{delta_rv_time_series}
\end{figure}

To visualise the different trends in the time series of the rest RV values, we computed the
rest RV differences for
four line combinations and list our results in Table \ref{line_combinations}; to this end,
we normalised the rest RV by subtracting the mean rest RV value and computed the
differences of these rest RV values. These RV differences ($\Delta$RV$_{norm\_rest}$) values are shown
in Fig. \ref{delta_rv_time_series}, which demonstrates 
a clear wave-like variation in the time range 2019.5-2020.5 during the great dimming.  Figure \ref{delta_rv_time_series}
also shows that the variations during the great dimming were
much stronger than those observed during the following season.  To quantify the strength of these
variations, the standard deviation of these
variations is computed for these two observation epochs and also for both individual seasons, with the results being
listed in Table \ref{line_combinations}.  

\begin{table}[!t]    
\caption{Standard deviations of the difference of the rest RV values for different line combinations}    
\label{line_combinations}    
\begin{center}    
\setlength{\tabcolsep}{5pt}
\scalebox{0.87}{
\begin{tabular}{cccc}
\hline    
\hline
                  & \multicolumn{3}{c}{$\sigma$ of $\Delta$ RV$_{norm\_rest}$ in different epochs} \\
Line combination & \footnotesize{2019.5-2021.5} & \footnotesize{2019.5-2020.5} & \footnotesize{2020.5-2021.5} \\ 
\hline
\noalign{\smallskip}
\ion{V}{I}$_{6251.82 \AA}$ - \ion{Fe}{I}$_{6252.56 \AA}$ & 1.31 & 2.43 & 0.83 \\
\ion{Fe}{I}$_{62521.56 \AA}$ - \ion{Fe}{I}$_{6254.26 \AA}$ & 1.53 & 2.49 & 0.55 \\
\ion{Fe}{I}$_{6254.26 \AA}$ - \ion{Ti}{I}$_{6261.10 \AA}$ & 1.93 & 2.70 & 1.09 \\
\ion{Ti}{I}$_{6261.10 \AA}$ - \ion{La}{II}$_{6262.30 \AA}$ & 2.62 & 4.29 & 0.93 \\
\noalign{\smallskip}\hline
\end{tabular}
}
\end{center}    
\end{table}

In summary, we exclude that the variations in the rest RV values are an
artifact created by an uncertain wavelength solution. Rather, we assume these variations to be a sign of additional
motions, suggesting that the behaviour of the variations in the rest RV values indicates a difference
in motion between the mean regions where the different lines are formed.

\subsection{Variation of the line depths and widths}

\begin{figure}
\centering
\includegraphics[scale=0.55]{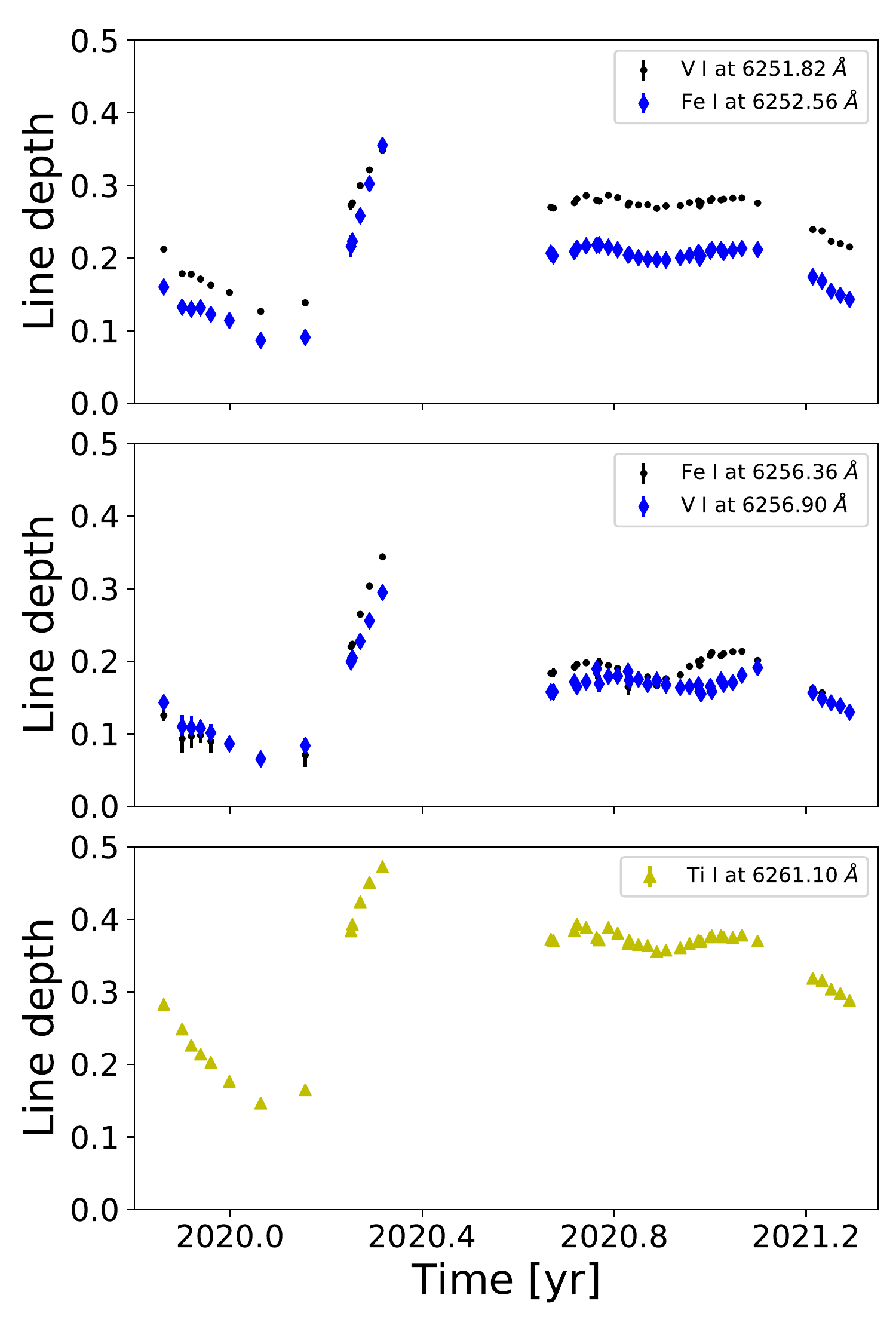}
\caption{Line depths vs. time shown.}
\label{ampl_time_series}
\end{figure}

\begin{figure}
\centering
\includegraphics[scale=0.55]{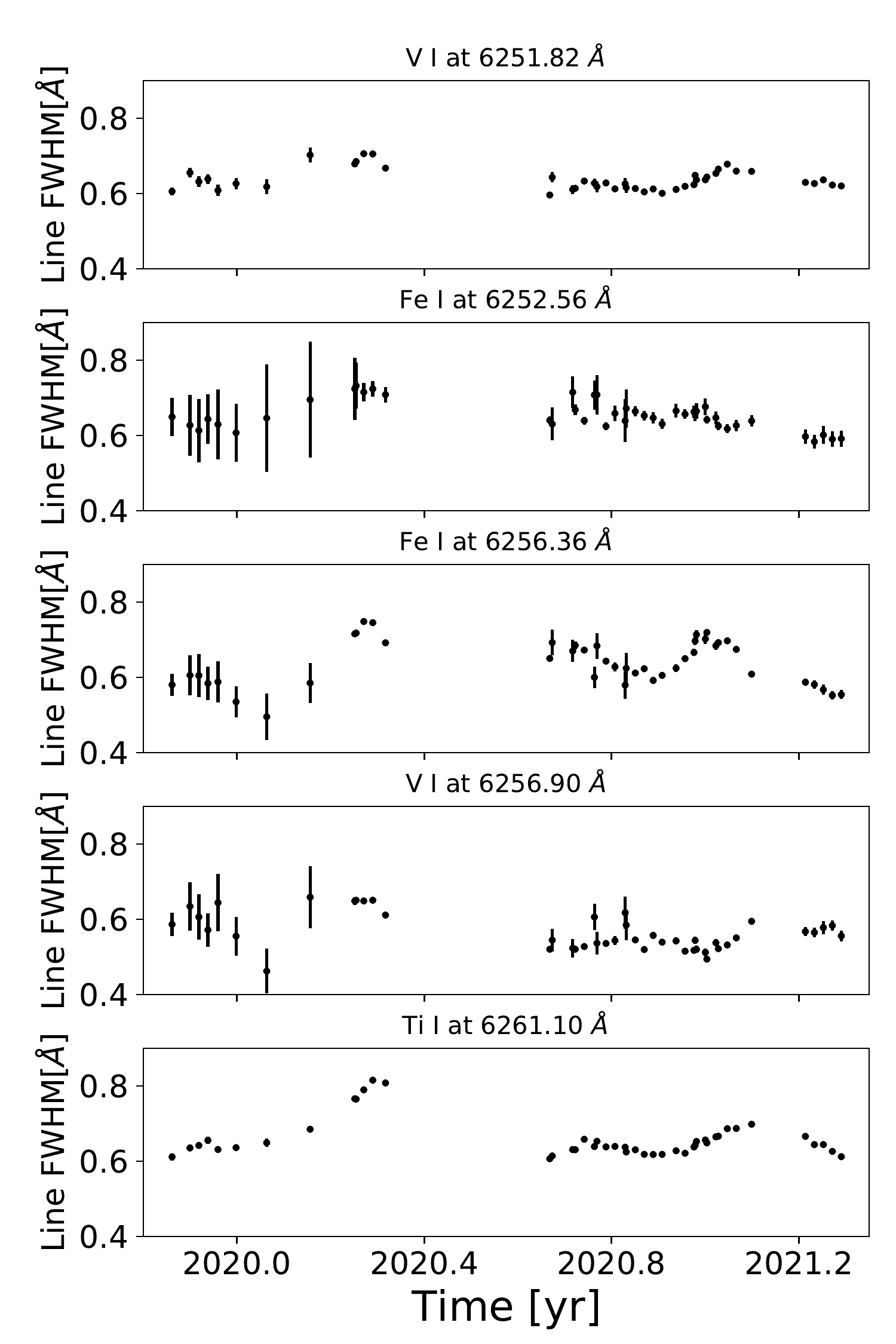}
\caption{Line FWHM vs. time shown.}
\label{FWHM_line_time_series}
\end{figure}

The strong line variability during the great dimming event is evident from
Fig.~\ref{spec_norm_example}. 
To visualise this significant variation in lines, we plot the line depths and the line widths
obtained from the Gaussian fits against observing epoch 
in Figs. \ref{ampl_time_series} and \ref{FWHM_line_time_series} for the lines of \ion{Fe}{I} at 6252.56 
and 6256.36 \AA, the \ion{V}{I} lines at 6251.82~\AA~and 6256.90 \AA, and for
the \ion{Ti}{I} lines at 6261.10 \AA. 

Both line parameters show clear variations with a wave-like shape reminiscent of
the effective temperature time series shown in Fig.\ref{tio_teff}. Furthermore, the strengths of the variations
in the line depths are about the same, while the strengths of the variations in the line widths are clearly different.
The differences in the variation of the line widths indicate differences in the extent of thermal broadening 
between the respective mean regions of the respective line origins.
Again, the variations in depth and width of the lines presented here suggest 
that the photosphere is always dynamic; however, during the great 
dimming event, the resulting changes were extraordinarily large.

\subsection{Temperature sensitive line-depth ratio of \ion{V}{I} 6251.82 \AA~over \ion{Fe}{I} 6252.56 \AA}
\label{ldr}

\citet{Gray2001PASP} demonstrated that the line-depth ratio (LDR) of the lines of  \ion{V}{I} at 6251.82~\AA~and
\ion{Fe}{I} at 6252.56~\AA~can be used to estimate the effective
 temperature for cool giants. The LDR is determined by the expression
\begin{eqnarray}\label{ldr_def}  
  LDR & = & \frac{line~depth~of~\ion{V}{I}_{~6251.82 \AA}}{line~depth~of~\ion{Fe}{I}_{~6252.56 \AA}}.
\end{eqnarray}

\citet{Gray2008AJ} used this LDR to study the variations of $\alpha$ Ori and showed it to
correlate with the V magnitude. However, 
\citet{Gray2008AJ} thought that a conversion of the LDR values into $\alpha$ 
Ori temperatures is not possible in absolute terms in the absence of an 
established relation. However, it is plausible that such a relation exists in 
physical terms, as we see below. 

\begin{figure}
\centering
\includegraphics[scale=0.5]{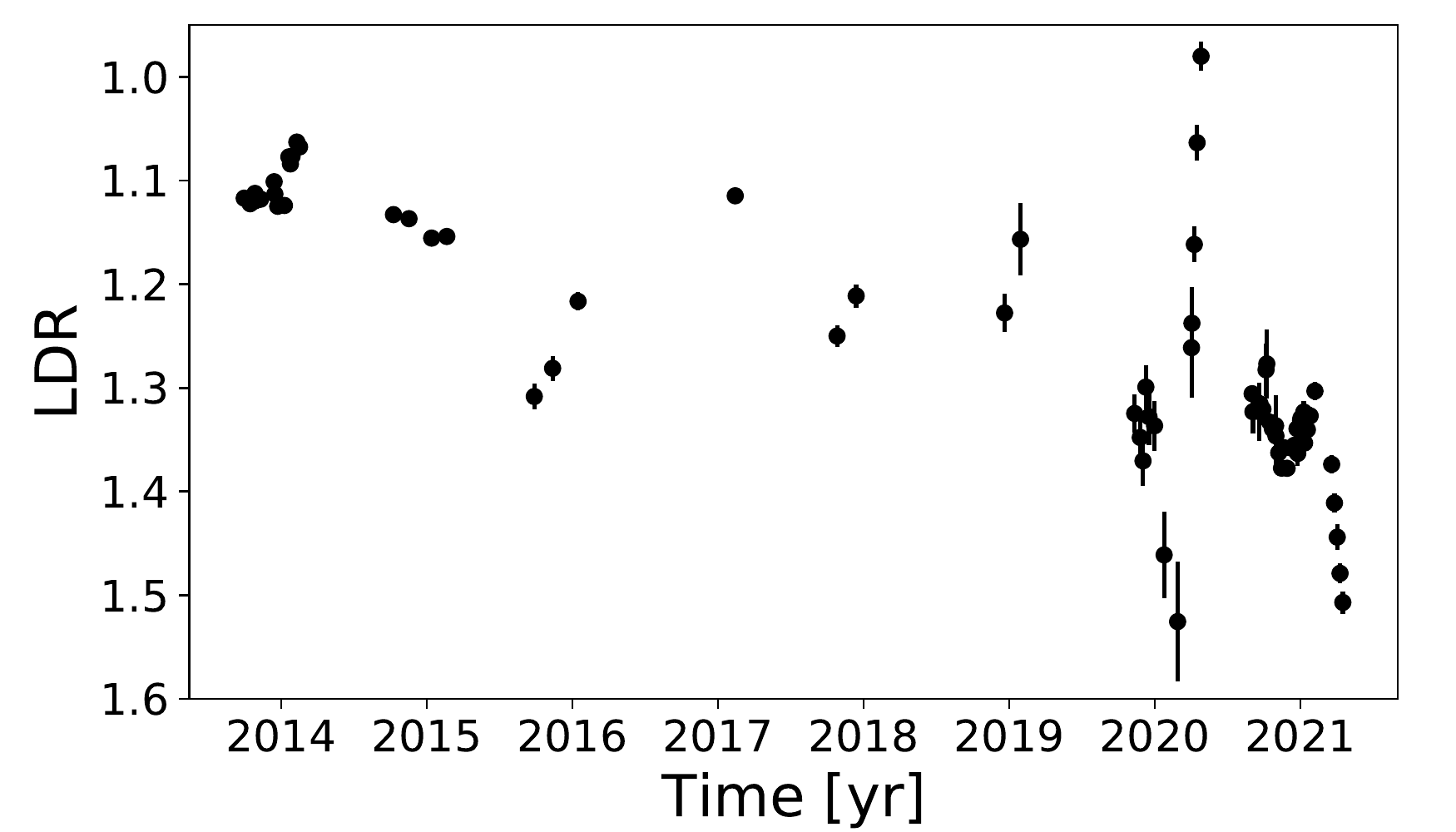}
\caption{Time series of the line-depth ratio of the \ion{V}{I} 6251.82 \AA~and 
        \ion{Fe}{I} 6252.56 \AA~lines}
\label{ldr_time_series}
\centering
\includegraphics[scale=0.5]{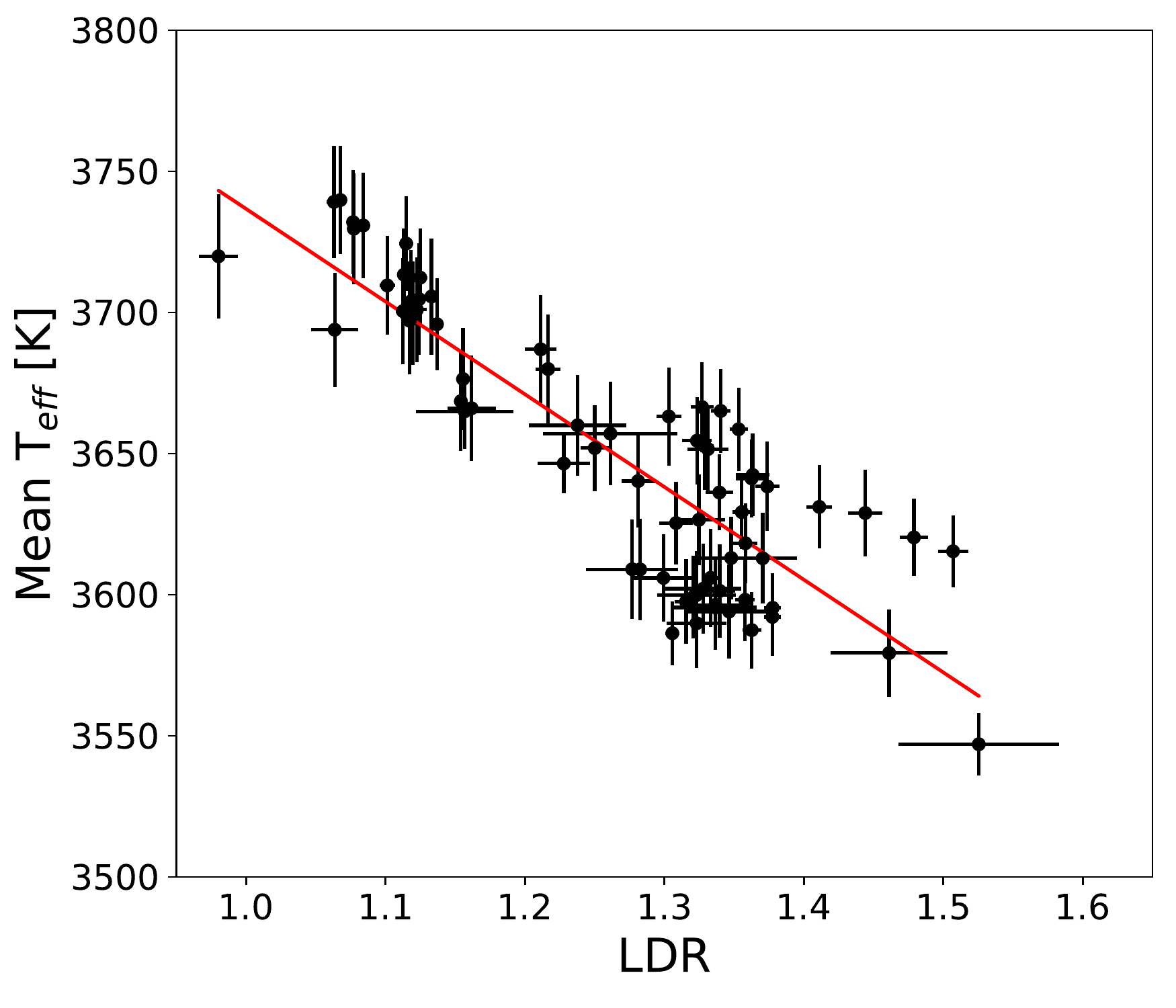}
\caption{T$_{\rm{eff}}$ over the line-depth ratio. The solid line 
        represents their mean relation.}
\label{ldr_teff}
\end{figure}

In this work, we measured the LDR of the \ion{V}{I} 6251.82 \AA~and \ion{Fe}{I} 
6252.56 \AA~lines to see whether any significant variations occurred during 
the great dimming event and how these compare to the normal variations of $\alpha$ Ori. 
For this purpose, we used the line depths obtained from the amplitudes of the 
Gaussian line fits above to compute the depth ratios of these lines; we show
our results in Fig. \ref{ldr_time_series}.  We see that
the LDR values do indeed vary with time and behave much like our derived effective 
temperature; see Sect. \ref{teff_tio_est}.

These results suggest a consistent relation between our effective temperature values
and the LDR (see Fig. \ref{ldr_teff}), with a simple linear behaviour, which is represented in
the same figure by the solid line:
\begin{eqnarray}\label{ldr_teff_rel}  
  T_{\rm{eff}} = (3736.7\pm7.8) K - (328.5\pm25.3) K (LDR-1).
\end{eqnarray}

The scatter of this relation is small: we obtain a standard deviation of the 
residuals of only 24 K.

\section{Summary and conclusions}

\begin{figure}
\centering
\includegraphics[scale=0.38]{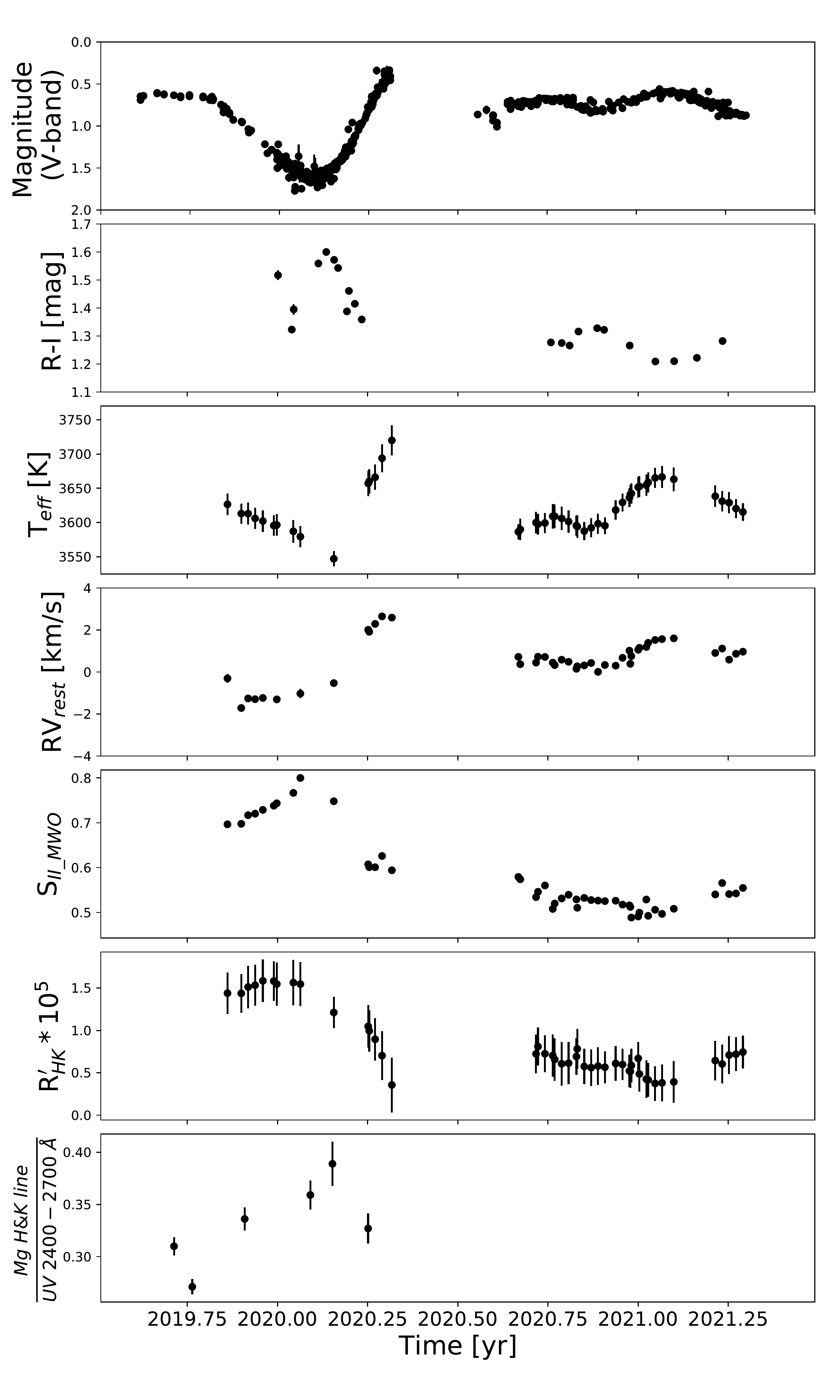}
\caption{Summary of the time series to show the main variation results. From top to bottom,
  we  show the V band, $R-I$ colour index, the effective temperature, the RV$_{rest}$ variation
  of the \ion{Ti}{I} line at 6261.1 \AA~and the TIGRE S$_{\rm{II\_MWO}}$,  the $R^{\prime}_{HK}$, and
  the ratio between the absolute \ion{Mg}{II} h\&k line flux and the MUV flux in 2400-2700~\AA~shown
  in \citet{Dupree2020ApJ}.}
\label{compare_results}
\end{figure}

In this study, we performed a multi-wavelength analysis of the variations observed
during the great dimming event of $\alpha$ Ori and in the subsequent
observation season of 2020 to 2021.  To provide an overview, we plot the derived
time series of the V magnitude, the well observed and temperature-dependent colour
index $R-I$, our effective 
temperature values, the RV$_{rest}$ km s$^{-1}$ of the line-core variation, 
our S-index time series, the derived $R^{\prime}_{HK}$ values, and the ratio between the 
absolute \ion{Mg}{II} h\&k line flux and the MUV flux in 2400-2700~\AA~shown in \citet{Dupree2020ApJ}
in Fig. \ref{compare_results};
as a reference of the long-term variability, the values of the V band magnitude are 
instructive and therefore, we show them here in the upper panel of Fig. \ref{compare_results}. 

Clearly, the values of effective temperature are important for a complete
understanding of the chromospheric activity of $\alpha$ Ori in absolute terms.
We analysed the variations in different colour indices using the photometric 
data of the AASVO, as well as of the LDR of the \ion{V}{I} 6251.82 \AA~and \ion{Fe}{I} 6252.56 \AA~lines;
as an example, we refer to the $R-I$ index time series in the third panel of 
Fig. \ref{compare_results}. This and the LDR time series do indicate changes in the
effective temperature, especially during the great dimming event.
To assess the effective temperature and its changes, we use four TiO bands covered by the TIGRE spectra 
and derive clear temperature variations of $\alpha$ Ori over time, in full agreement 
with \citet{Harper2020ApJ}. 

Next, we study the magnetic activity of $\alpha$ Ori, starting with the \ion{Ca}{II} K line
during the great dimming event. We find a strengthening 
of the chromospheric line-core emission relative to the photospheric spectrum 
in the months leading up to the brightness minimum, and afterwards 
a weakening of the relative \ion{Ca}{II} K line-core brightness. 
We then compare the wider 2{\AA} S-indices taken with TIGRE with the respective
values obtained by the Mount Wilson program in the 1980s for giants, including 
$\alpha$ Ori. To this end, we needed to derive a transformation equation between those 
two scales based on a larger number of giants observed by the present authors and Mount Wilson
in order to average out most individual activity changes between those two epochs.  
Our analysis then shows that
the current TIGRE measurements of the Ca~II lines can be related to
Mount Wilson S-index data taken much earlier and to the photometric data available 
from the AAVSO database over the past four decades.
We used these relations to construct a long-term light curve of  $\alpha$ Ori that extends
over almost 40 years,
thus providing a very much extended perspective of the great dimming event.

Upon subsequent inspection of the combined 2{\AA} S-index time series, we note three larger increases in
the S-values,
each lasting less than a year, occurring  in 1984-1985, 1989 (both Mount Wilson data), and 2020 
(TIGRE); the latter coincides with the great dimming event and is shown here 
in the fifth panel of Fig. \ref{compare_results}.  A comparison with the V-band light curve
shows a clear  brightness decrease  over the same period, suggesting
a correlation between the decrease in brightness and the increase in the S-values.
This correlation indicates that the debated former dimming in 1984 and 1985 was a real event
---even though only two observers recorded this event in the visual band---
because it coincided with the first 2{\AA} S-increase observed by Mount Wilson.

Next, we estimated the \ion{Ca}{II} H\&K line flux in absolute terms, applying the 
method by \citet{Linsky1979ApJS} to $\alpha$ Ori. The time series of the 
absolute \ion{Ca}{II} H\&K line flux obtained in this way shows, in stark contrast to the S-value time series,
that at the beginning of the great dimming event there was clearly a larger chromospheric
emission than at the brightness minimum. The time series rather follows the shape of the effective
temperature graph during the great dimming, which is no surprise, because the continuum 
flux ---used as a reference to estimate the line flux--- depends very strongly on the
effective temperature in this NUV spectral range.  Ignoring this effect would be
like comparing the fluxes of two different stars, because the effective
temperature of $\alpha$ Ori  changes  notably during the great dimming event. 

We also estimated the excess of the normalised \ion{Ca}{II} H\&K line
flux, $R^{\prime}_{HK}$.  Interestingly, this excess flux does not change significantly at the
beginning of the great dimming event (see sixth panel of Fig. \ref{compare_results}), 
and only starts to vary  after the dimming event.  Also,  
comparing the absolute \ion{Ca}{II} H\&K line flux with the absolute \ion{Mg}{II} 
h\&k line flux, our results  and those of \citet{Dupree2020ApJ}  are comparable. 
We therefore
conclude  that in order to remove the temperature effect on the latter, we need to involve
the ratio between the absolute \ion{Mg}{II} h\&k line flux and the MUV flux in 2400-2700~\AA~ 
(shown in \citet{Dupree2020ApJ};  see Figs. \ref{mg_II_hk_flux_dupree}
and the seventh panel of Fig. \ref{compare_results}). This ratio shows
a behaviour that is comparable to that seen in our $R^{\prime}_{HK}$ time series during the great dimming.

This finding implies that the chromospheric activity did not change much
at the beginning of the observed brightness minimum; however, as seen in the time series 
of  $R^{\prime}_{HK}$, there is a decrease in chromospheric activity
after the brightness minimum in February 2020. Therefore, our interpretation contradicts
that of \citet{Dupree2020ApJ}, at least as far
as the cause of the dimming is concerned. Whatever started this chain of events does
not seem to be located in the outer atmosphere.

This conclusion is supported by the observed variations in different lines, which
display strong changes during the great dimming event, both in their ratio of line strength, which are caused by the impact of the 
effective temperature changes, and also in their width and
respective RV; an example is shown in the fourth panel of Fig. \ref{compare_results}.
This dynamical behaviour indicates a strong variation in the photosphere of $\alpha$ Ori.

In agreement with previous studies, our results 
demonstrate that $\alpha$ Ori is a highly variable star, which is reflected in all parameters
illustrated in Fig. \ref{compare_results}. 
This leads us to the interesting question of how these variations provide support to 
the two main hypotheses for the cause of the great dimming event.
The decrease in the mean effective temperature, as demonstrated here, in
combination with a reduction in the effective luminous area of the photosphere
 ---which may not necessarily be a radius change--- supports
scenarios (see \citet{Freytag2019IAUS..343....9F})
where a large cool area forms temporarily, while a much smaller 
number of warm convection cells rise. Such temporary quasi-synchronisation
of several large convection cells may go hand in hand with shock- or density waves. 
The down-flow of such a large, temporary  cool area, which greatly  reduces
both the effective luminous area of the photosphere and the effective 
temperature, would not appear as obvious  changes in RV; cooler regions are 
under-represented in the observed line profiles, because cooler gas contributes
much less to the total emergent flux, and we would rather expect changes on a differential
scale between lines with different formation temperatures, as reported here.
Thus, all these ideas point to the photosphere 
and its convection as the main driver of the great dimming event.

The other prominent hypothesis, put forward by \citet{Dupree2020ApJ}, 
assumes veiling by dust related to an observed mass ejection in 
October 2019. This idea is also based 
on the stronger absolute \ion{Mg}{II} h\&k line flux and MUV flux in 2400-2700 \AA~seen
in October 2019 compared to the respective fluxes during the brightness 
minimum in February 2020 reported by the same authors. In our study we too find 
that the absolute \ion{Ca}{II} H\&K line flux was clearly larger in November 2019 
than during the brightness minimum in February 2020. However, the significant  changes in effective
temperature and luminous area must be accounted for in order to correctly
assess the chromospheric energy output on an absolute scale as shown above.
As a consequence, the apparent large chromospheric emission increase that was observed to take place simultaneously with 
the beginning of the great dimming event disappears,  and it is not until after the 
bolometric brightness minimum that the chromosphere reacts to it in absolute terms.

Thus, the corrected timing of the different events disagrees with the assumption of a strong mass ejection 
in October 2019, and any subsequent veiling by dust is therefore an unlikely cause of 
the great dimming;  it would rather be a consequence and serve as an enhancement. 
Further  evidence in favour of an origin of the chain of events in the photosphere 
and its convection, forming sometimes larger cool plumes, is provided by the 
recurrence of such
an event; we show the similarities between the available data from
1984-1985 and those from 2020-2021. Given the extremely large size and therefore small number of 
convection cells under the extremely low gravity of $\alpha$ Ori, any temporary 
chance synchronisation and plume formation on the earth-facing side should 
indeed not be a particularly unique event.

\begin{acknowledgements}  
We are grateful to our anonymous referee for the kind report.--
We particularly wish to thank our colleague Bernd Freytag of 
the Theoretical Astrophysics Department, Uppsala University, for 
enlightening discussions of how to interpret the complex information 
gained from our observations. --
We particularly acknowledge use of the public HK\_Project\_v1995\_NSO data base of 
the Mount Wilson Observatory HK Project, 
which was supported by both public and private funding through the Carnegie 
Observatories, the Mount Wilson Institute, and the Harvard-Smithsonian Center 
for Astrophysics, running from 1960ies over more than three decades.
These data are the result of the dedicated work of O. Wilson, A.
Vaughan, G. Preston, D. Duncan, S. Baliunas, and many others.
We are also very grateful for using the AAVSO International Database, 
photometric data on $\alpha$ Ori contributed by observers worldwide. --
Furthermore, the VizieR catalogue access tool, CDS, Strasbourg, France 
(see the original description of the VizieR service: A\&AS 143, 23) was
helpful to this research, as well as the SVO Filter Profile Service (http://svo2.cab.inta-csic.es/theory/fps/)
supported from the Spanish MINECO through grant AYA2017-84089. --
Finally, our research and international collaboration between Germany and Mexico, as 
well as the operation of TIGRE, has benefited significantly from travel money by the 
Conacyt-DFG bilateral projects No. 207772 and 278156, and by the institutional 
support given by the universities of Hamburg and Guanajuato (UG: by
DAIP funding from its project 003/2022 and CAP programs) over the past decade.
\end{acknowledgements}  
  
\bibliographystyle{aa}  
\bibliography{bibfile}

\begin{thebibliography}{37}
\expandafter\ifx\csname natexlab\endcsname\relax\def\natexlab#1{#1}\fi

\bibitem[{{Choi} {et~al.}(1995){Choi}, {Soon}, {Donahue}, {Baliunas}, \&
  {Henry}}]{choi1995PASP}
{Choi}, H.-J., {Soon}, W., {Donahue}, R.~A., {Baliunas}, S.~L., \& {Henry},
  G.~W. 1995, \pasp, 107, 744

\bibitem[{{Dharmawardena} {et~al.}(2020){Dharmawardena}, {Mairs}, {Scicluna},
  {Bell}, {McDonald}, {Menten}, {Weiss}, \& {Zijlstra}}]{Dharmawardena2020ApJ}
{Dharmawardena}, T.~E., {Mairs}, S., {Scicluna}, P., {et~al.} 2020, \apjl, 897,
  L9

\bibitem[{{Duncan} {et~al.}(1991){Duncan}, {Vaughan}, {Wilson}, {Preston},
  {Frazer}, {Lanning}, {Misch}, {Mueller}, {Soyumer}, {Woodard}, {Baliunas},
  {Noyes}, {Hartmann}, {Porter}, {Zwaan}, {Middelkoop}, {Rutten}, \&
  {Mihalas}}]{duncan1991}
{Duncan}, D.~K., {Vaughan}, A.~H., {Wilson}, O.~C., {et~al.} 1991, \apjs, 76,
  383

\bibitem[{{Dupree} {et~al.}(2020){Dupree}, {Strassmeier}, {Matthews},
  {Uitenbroek}, {Calderwood}, {Granzer}, {Guinan}, {Leike}, {Montarg{\`e}s},
  {Richards}, {Wasatonic}, \& {Weber}}]{Dupree2020ApJ}
{Dupree}, A.~K., {Strassmeier}, K.~G., {Matthews}, L.~D., {et~al.} 2020, \apj,
  899, 68

\bibitem[{{ESA}(1997)}]{HIPPARCOS1997ESA}
{ESA}, ed. 1997, ESA Special Publication, Vol. 1200, {The HIPPARCOS and TYCHO
  catalogues. Astrometric and photometric star catalogues derived from the ESA
  HIPPARCOS Space Astrometry Mission}

\bibitem[{{Freytag} {et~al.}(2019){Freytag}, {H{\"o}fner}, \&
  {Liljegren}}]{Freytag2019IAUS..343....9F}
{Freytag}, B., {H{\"o}fner}, S., \& {Liljegren}, S. 2019, IAU Symposium, 343, 9

\bibitem[{{Gonz{\'a}lez-P{\'e}rez} {et~al.}(2022){Gonz{\'a}lez-P{\'e}rez},
  {Mittag}, {Schmitt}, {Schr{\"o}der}, {Jack}, {Rauw}, \&
  {Naz{\'e}}}]{Gonzalez-Perez_2022}
{Gonz{\'a}lez-P{\'e}rez}, J.~N., {Mittag}, M., {Schmitt}, J. H.~M.~M., {et~al.}
  2022, Frontiers in Astronomy and Space Sciences, 9, 912546

\bibitem[{{Gray}(2005)}]{Gray2005oasp.book}
{Gray}, D.~F. 2005, {The Observation and Analysis of Stellar Photospheres}

\bibitem[{{Gray}(2008)}]{Gray2008AJ}
{Gray}, D.~F. 2008, \aj, 135, 1450

\bibitem[{{Gray} \& {Brown}(2001)}]{Gray2001PASP}
{Gray}, D.~F. \& {Brown}, K. 2001, \pasp, 113, 723

\bibitem[{{Guinan} {et~al.}(2019){Guinan}, {Wasatonic}, \&
  {Calderwood}}]{Guinan2019ATel13341}
{Guinan}, E.~F., {Wasatonic}, R.~J., \& {Calderwood}, T.~J. 2019, The
  Astronomer's Telegram, 13341, 1

\bibitem[{{Hall} \& {Lockwood}(1995)}]{Hall1995ApJ}
{Hall}, J.~C. \& {Lockwood}, G.~W. 1995, \apj, 438, 404

\bibitem[{{Harper} {et~al.}(2020){Harper}, {Guinan}, {Wasatonic}, \&
  {Ryde}}]{Harper2020ApJ}
{Harper}, G.~M., {Guinan}, E.~F., {Wasatonic}, R., \& {Ryde}, N. 2020, \apj,
  905, 34

\bibitem[{Hauschildt {et~al.}(1999)Hauschildt, Allard, \&
  Baron}]{hauschildt1999}
Hauschildt, P.~H., Allard, F., \& Baron, E. 1999, ApJ, 512, 377

\bibitem[{{Hempelmann} {et~al.}(2016){Hempelmann}, {Mittag}, {Gonzalez-Perez},
  {Schmitt}, {Schr{\"o}der}, \& {Rauw}}]{Hempelmann2016}
{Hempelmann}, A., {Mittag}, M., {Gonzalez-Perez}, J.~N., {et~al.} 2016, \aap,
  586, A14

\bibitem[{{Husser} {et~al.}(2013){Husser}, {Wende-von Berg}, {Dreizler},
  {Homeier}, {Reiners}, {Barman}, \& {Hauschildt}}]{Husser2013A&A}
{Husser}, T.~O., {Wende-von Berg}, S., {Dreizler}, S., {et~al.} 2013, \aap,
  553, A6

\bibitem[{{Joyce} {et~al.}(2020){Joyce}, {Leung}, {Moln{\'a}r}, {Ireland},
  {Kobayashi}, \& {Nomoto}}]{Joyce2020ApJ}
{Joyce}, M., {Leung}, S.-C., {Moln{\'a}r}, L., {et~al.} 2020, \apj, 902, 63

\bibitem[{{Kafka}(2021)}]{Kafka}
{Kafka}, S. 2021, {Observations from the AAVSO International Database,
  https://www.aavso.org}

\bibitem[{{Kharchenko} {et~al.}(2007){Kharchenko}, {Scholz}, {Piskunov},
  {R{\"o}ser}, \& {Schilbach}}]{Kharchenko2007AN}
{Kharchenko}, N.~V., {Scholz}, R.~D., {Piskunov}, A.~E., {R{\"o}ser}, S., \&
  {Schilbach}, E. 2007, Astronomische Nachrichten, 328, 889

\bibitem[{Kramida {et~al.}(2021)Kramida, {Yu.~Ralchenko}, Reader, \& {and NIST
  ASD Team}}]{NIST_ASD}
Kramida, A., {Yu.~Ralchenko}, Reader, J., \& {and NIST ASD Team}. 2021, {NIST
  Atomic Spectra Database (ver. 5.9), [Online]. Available:
  {\tt{https://physics.nist.gov/asd}} [2021, December 29]. National Institute
  of Standards and Technology, Gaithersburg, MD.}

\bibitem[{{Lan{\c{c}}on} {et~al.}(2007){Lan{\c{c}}on}, {Hauschildt}, {Ladjal},
  \& {Mouhcine}}]{lancon2007A&A}
{Lan{\c{c}}on}, A., {Hauschildt}, P.~H., {Ladjal}, D., \& {Mouhcine}, M. 2007,
  \aap, 468, 205

\bibitem[{{Levesque} \& {Massey}(2020)}]{Levesque2020ApJ}
{Levesque}, E.~M. \& {Massey}, P. 2020, \apjl, 891, L37

\bibitem[{{Linsky} {et~al.}(1979){Linsky}, {Worden}, {McClintock}, \&
  {Robertson}}]{Linsky1979ApJS}
{Linsky}, J.~L., {Worden}, S.~P., {McClintock}, W., \& {Robertson}, R.~M. 1979,
  \apjs, 41, 47

\bibitem[{{Mann} \& {von Braun}(2015)}]{Mann2015PASP}
{Mann}, A.~W. \& {von Braun}, K. 2015, \pasp, 127, 102

\bibitem[{{Mittag} {et~al.}(2010){Mittag}, {Hempelmann},
  {Gonz{\'a}lez-P{\'e}rez}, \& {Schmitt}}]{mittag2010}
{Mittag}, M., {Hempelmann}, A., {Gonz{\'a}lez-P{\'e}rez}, J.~N., \& {Schmitt},
  J.~H.~M.~M. 2010, Advances in Astronomy, 2010, 101502

\bibitem[{{Mittag} {et~al.}(2016){Mittag}, {Schr{\"o}der}, {Hempelmann},
  {Gonz{\'a}lez-P{\'e}rez}, \& {Schmitt}}]{Mittag2016A&A}
{Mittag}, M., {Schr{\"o}der}, K.~P., {Hempelmann}, A.,
  {Gonz{\'a}lez-P{\'e}rez}, J.~N., \& {Schmitt}, J.~H.~M.~M. 2016, \aap, 591,
  A89

\bibitem[{{Montarg{\`e}s} {et~al.}(2021){Montarg{\`e}s}, {Cannon}, {Lagadec},
  {de Koter}, {Kervella}, {Sanchez-Bermudez}, {Paladini}, {Cantalloube},
  {Decin}, {Scicluna}, {Kravchenko}, {Dupree}, {Ridgway}, {Wittkowski},
  {Anugu}, {Norris}, {Rau}, {Perrin}, {Chiavassa}, {Kraus}, {Monnier},
  {Millour}, {Le Bouquin}, {Haubois}, {Lopez}, {Stee}, \&
  {Danchi}}]{Montarges2021Natur}
{Montarg{\`e}s}, M., {Cannon}, E., {Lagadec}, E., {et~al.} 2021, \nat, 594, 365

\bibitem[{{Piskunov} \& {Valenti}(2002)}]{REDUCE2002A&A385.1095P}
{Piskunov}, N.~E. \& {Valenti}, J.~A. 2002, \aap, 385, 1095

\bibitem[{Radick \& Pevtsov(2018)}]{Radick_MWO}
Radick, R. \& Pevtsov, A. 2018, in HK\_Project\_v1995\_NSO (Harvard Dataverse)

\bibitem[{{Rodrigo} \& {Solano}(2020)}]{Rodrigo2020sea..confE.182R}
{Rodrigo}, C. \& {Solano}, E. 2020, in Contributions to the XIV.0 Scientific
  Meeting (virtual) of the Spanish Astronomical Society, 182

\bibitem[{{Rodrigo} {et~al.}(2012){Rodrigo}, {Solano}, \&
  {Bayo}}]{Rodrigo2012ivoa.rept.1015R}
{Rodrigo}, C., {Solano}, E., \& {Bayo}, A. 2012, {SVO Filter Profile Service
  Version 1.0}, IVOA Working Draft 15 October 2012

\bibitem[{{Rutten}(1984)}]{Rutten1984A&A}
{Rutten}, R.~G.~M. 1984, \aap, 130, 353

\bibitem[{{Schmitt} {et~al.}(2014){Schmitt}, {Schr{\"o}der}, {Rauw},
  {Hempelmann}, {Mittag}, {Gonz{\'a}lez-P{\'e}rez}, {Czesla}, {Wolter}, {Jack},
  {Eenens}, \& {Trinidad}}]{Schmitt2014AN335787S}
{Schmitt}, J.~H.~M.~M., {Schr{\"o}der}, K.-P., {Rauw}, G., {et~al.} 2014,
  Astronomische Nachrichten, 335, 787

\bibitem[{{Soubiran} {et~al.}(2016){Soubiran}, {Le Campion}, {Brouillet}, \&
  {Chemin}}]{Soubiran2016}
{Soubiran}, C., {Le Campion}, J.-F., {Brouillet}, N., \& {Chemin}, L. 2016,
  \aap, 591, A118

\bibitem[{{Vaughan} {et~al.}(1978){Vaughan}, {Preston}, \&
  {Wilson}}]{Vaughan1978PASP90267V}
{Vaughan}, A.~H., {Preston}, G.~W., \& {Wilson}, O.~C. 1978, \pasp, 90, 267

\bibitem[{{Vernazza} {et~al.}(1981){Vernazza}, {Avrett}, \&
  {Loeser}}]{Vernazza1981ApJS}
{Vernazza}, J.~E., {Avrett}, E.~H., \& {Loeser}, R. 1981, \apjs, 45, 635

\bibitem[{{Wilson}(1982)}]{Wilson1982ApJ}
{Wilson}, O.~C. 1982, \apj, 257, 179

\end{thebibliography}

\begin{appendix}
\section{Variation of the calculated radius and calculated $\log(g)$}
\label{var_stellar_radius}
\citet{Dupree2020ApJ} presented RV measurements of $\alpha$ Ori 
during the great dimming event. These were variable and are correlated with 
the observed V mag variation. If there are real radius
changes, the question therefore arises as to which changes we address here in terms of the calculated radius. Indeed, the photometric
variation cannot be explained by the effective temperature changes alone and implies
changes in the calculated radius. However, we cannot distinguish whether these reflect physical radius changes or simply
changes in the effective luminous area. A reduced calculated
radius could also result from the temporary presence of a large dark patch on
the photosphere, such as that seen when there is a rising or sinking cooled plume formed in the course of some chance
synchronisation of the few large granules on the observable side of this supergiant.

To estimate such changes in the stellar radius of $\alpha$ Ori in absolute terms, 
the luminosity of $\alpha$ Ori can be used: assuming spherical symmetry, it is
proportional to $R^{2}T_{eff}^{4}$. For a quantification of the stellar luminosity 
of $\alpha$ Ori, we adopted the distance of 168 pc given by \citep{Joyce2020ApJ}. 
In addition, we used the V magnitude data from the AAVSO database of the days where we 
were able to derive an effective temperature.
 
\begin{figure}
\centering
\includegraphics[scale=0.4]{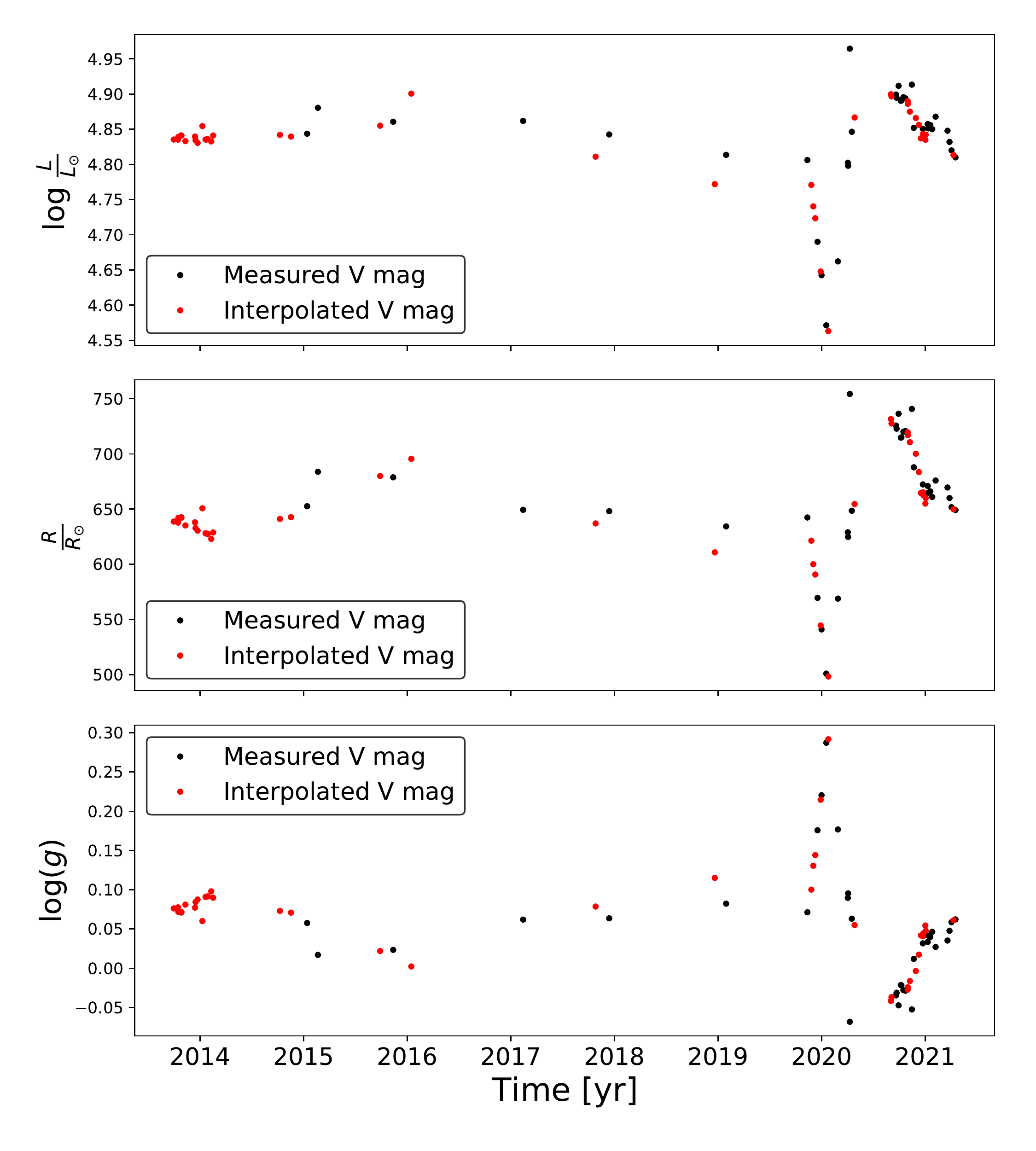}
\caption{Calculated stellar properties where BC is constant vs. time: Upper panel: Logarithmic luminosity in solar units vs. time. 
        Middle and lower panels: Calculated radius in solar units and calculated gravity 
        (calculated $\log(g)$) on the same timescales. Here, a constant BC of -1.93 was used, which is
        based on the long-term mean effective temperature.}
\label{logL_radius_log}
\end{figure}

For days with an effective temperature value but no V mag, we interpolated the nearest
V mag data. The results of this calculation are shown in the upper panel of Fig \ref{logL_radius_log}. 
For simplicity, the bolometric correction (BC) of -1.93 here is based on the long-term 
mean effective temperature of 3649 K, but see below. For the calculation of the 
latter, we apply Eq. 10.10 of \citet[]{Gray2005oasp.book}.

The strong decrease in the stellar luminosity during the great dimming event is
obvious, and radius and gravity values calculated from luminosity and effective temperature are
presented in the middle and lower panels of Fig. \ref{logL_radius_log}. As the 
luminosity drop was too large to be explained by the lower effective temperature alone,
the radius would had to decrease from $\approx$640 $R_{\odot}$ on 11~November~2019 to 
$\approx$500 $R_{\odot}$ on 24 January 2020, which would explain the above-mentioned
variation of the RV data.

The $\log(g)$ values were then calculated using the above stellar radius values 
and assuming a (constant) mean stellar mass of 17.75 $M_{\odot}$ 
(see \citet{Joyce2020ApJ}). As seen in the lower panel of Fig. \ref{logL_radius_log},
there is an increase in the calculated $\log(g)$ of up to $\approx$0.3 dex during
the great dimming event, which is caused by the reduction in calculated radius.

\begin{figure}
\centering
\includegraphics[scale=0.40]{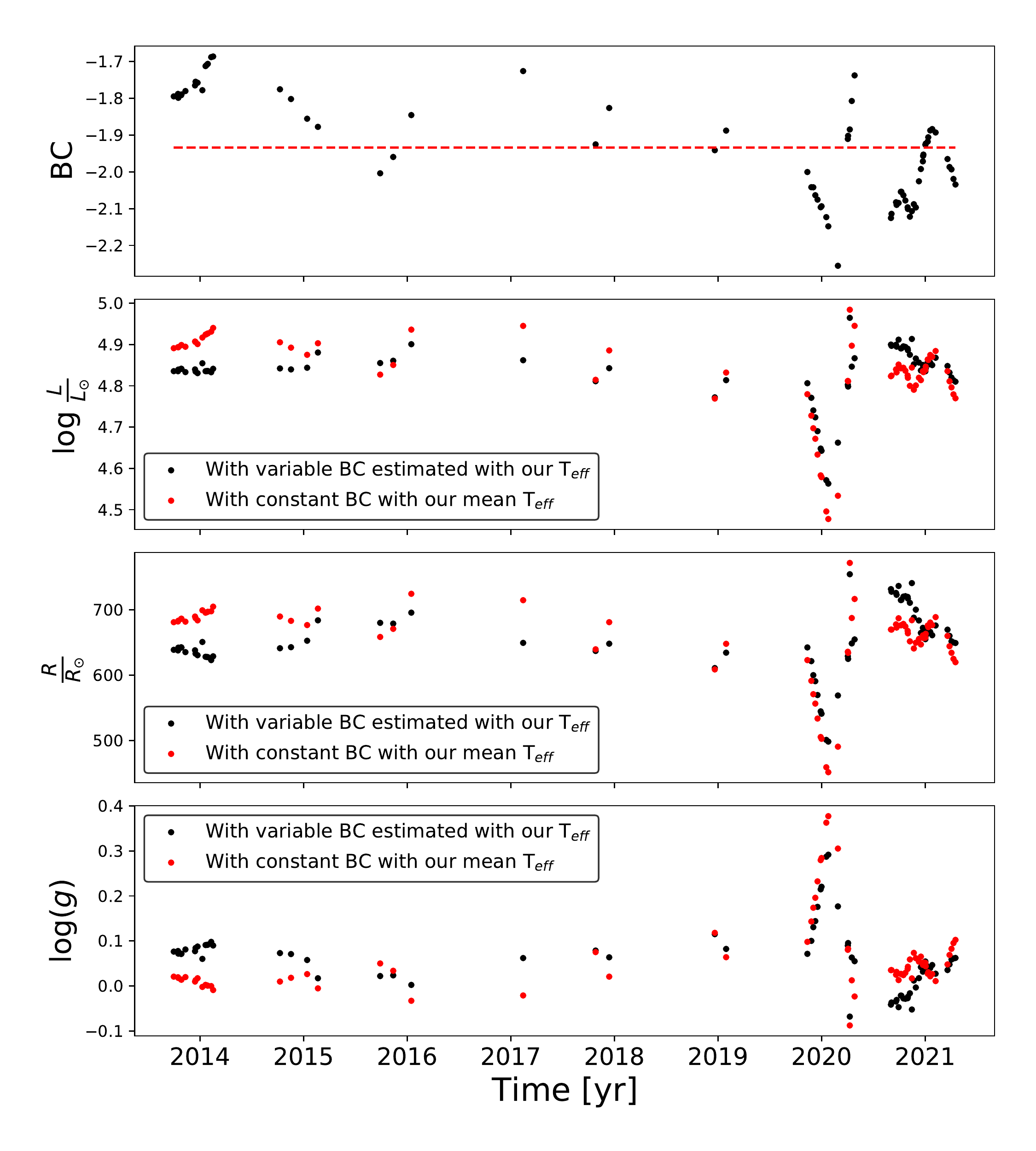}
\caption{Calculated Stellar properties where BC is variable vs. time: Upper panel: Time-variable BC, using the derived daily
effective temperatures (black dots). The red dashed line shows the BC of the mean effective temperature of 3649 K.
  Second panel:  Logarithmic luminosity. Third panel:
  Stellar radius. Lowest panel: $\log(g)$. 
  Black points represent the values obtained with daily,
  time-variable BC (as in the top panel), and red points show the values using the constant BC
  instead.}
\label{compare_logL_radius_logg_with_const_BC}
\end{figure}

In this context, the impact of the temperature variation on the bolometric correction 
(BC) now requires some consideration. For this purpose, we calculated the BCs
for the 
individual days (see top panel of Fig. \ref{compare_logL_radius_logg_with_const_BC})
where we had derived the effective temperature, 
again using the relation in Eq 10.10 of \citet[]{Gray2005oasp.book}.
We repeated the calculation of the luminosity, radius, and gravity to assess the 
impact of a time-variable BC (in the range of -1.69 to -2.26 mag) and compared the results
with those obtained assuming a constant BC (-1.93) based on the long-term mean effective temperature 
of 3649 K.
These comparisons are shown in Fig. \ref{compare_logL_radius_logg_with_const_BC}.

The second panel of Fig. \ref{compare_logL_radius_logg_with_const_BC} shows the
logarithm of the luminosity in solar units versus time. the third panel depicts the 
calculated radius, and the lowest panel the calculated $\log(g)$ (if the calculated radius was indeed
representative of the geometrical radius). In all panels, black points represent
the values calculated with the time-variable BC based on the daily effective 
temperatures, and red points depict the values for which the constant BC of the long-term 
mean effective temperature is used.

Not surprisingly, we find some differences between the use of the daily time-variable 
BC and a constant mean effective temperature BC for the stellar luminosity,
calculated radius, and calculated $\log(g)$. In principle, these differences vary with the course
of the effective temperatures and amplify the effect of the latter on all three quantities.
From this test, we conclude that all the consequences of the effective temperature changes must be considered 
. 

\section{Theoretical colour estimation using PHOENIX model spectra}
\label{phx_colour}
To understand, at least qualitatively, the non-linear behaviour of the different
colour indices $B-V$, $V-R$, $R-I,$ and $J-H$ mag with a variable effective
temperature in the range from 3300~K to 4000~K, we estimated these indices
from synthetic PHOENIX model spectra of different temperatures. 
Here we give an example to our work,
using the synthetic spectrum of the model atmosphere for T$_{eff}$ = 3600 K, 
log(g) = 0, [Fe/H] = 0, as made available by the University of G\"ottingen 
database of \citep{Husser2013A&A}.

\begin{figure}
\centering
\includegraphics[scale=0.38]{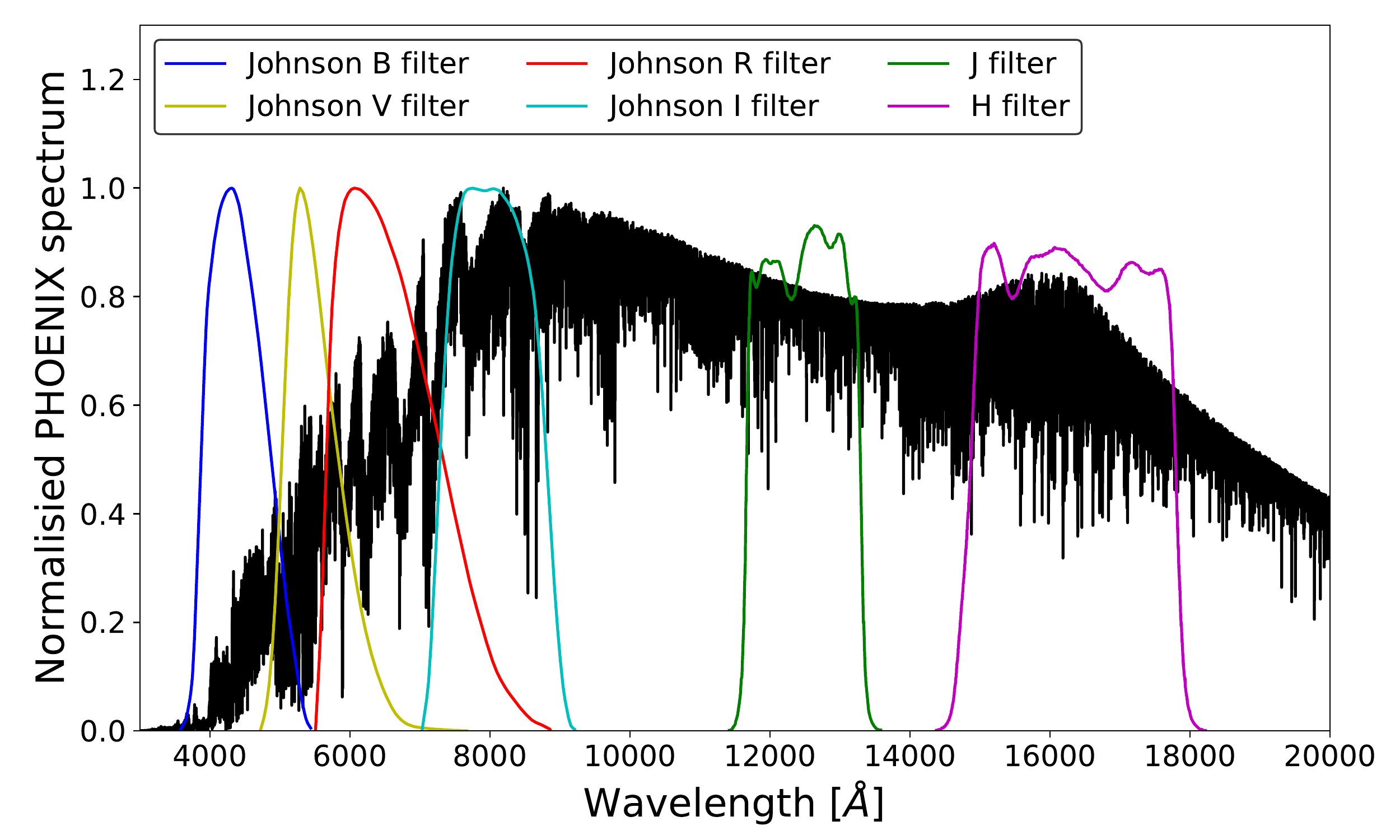}
\caption{Normalised PHOENIX spectrum (T$_{eff}$ = 3600 K, log(g) = 0 and [Fe/H] = 0),;
  over-plotted are the filter functions used here.}
\label{phx_spectrum_and_filter_function}
\end{figure}

\begin{figure}
\centering
\includegraphics[scale=0.55]{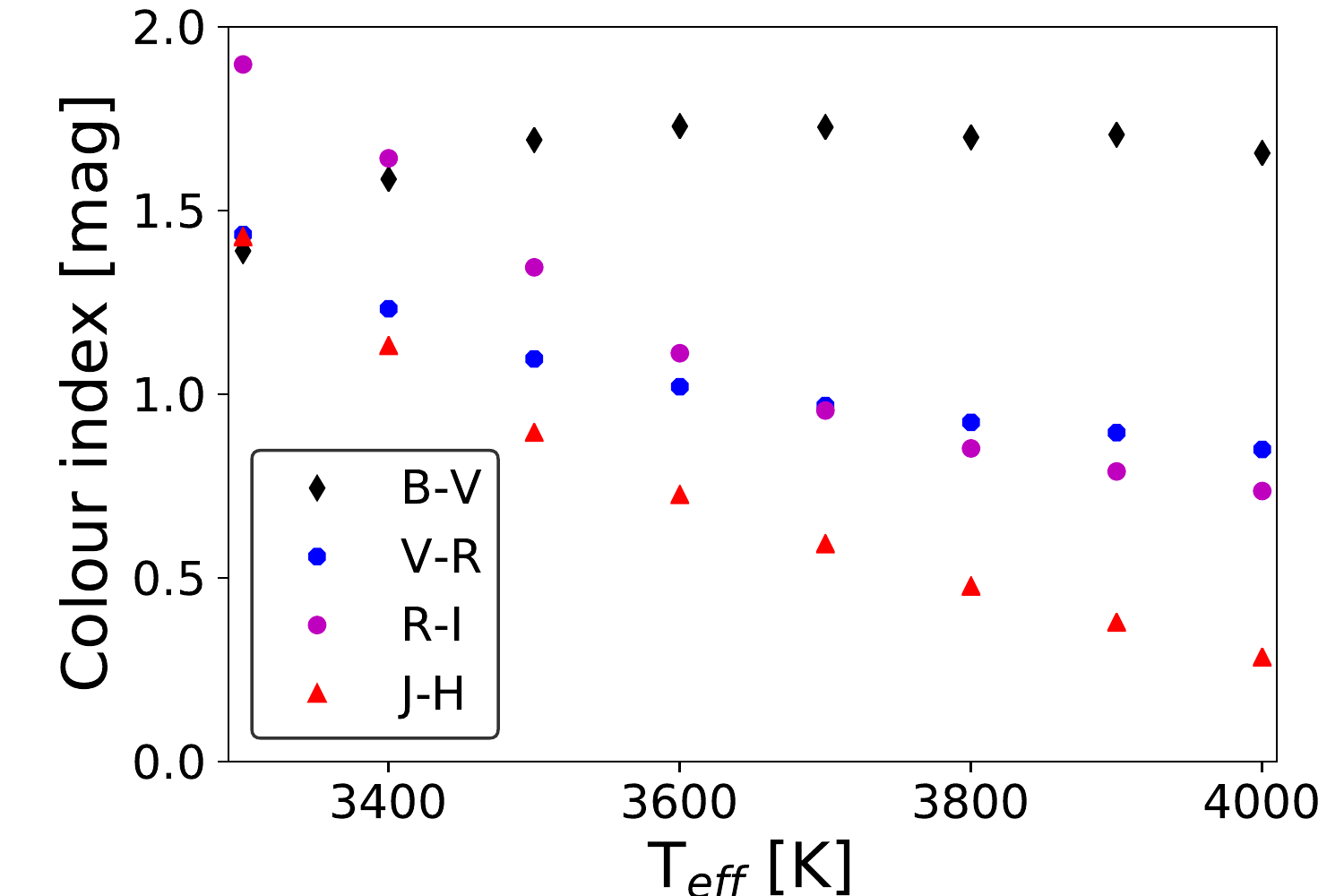}
\caption{$B-V$, $V-R$, $R-I,$ and $J-H$ colour indices vs. effective temperature.}
\label{phx_colour_indices}
\end{figure}

The filter functions for the B, V, R, and I bands were taken from \citet{Mann2015PASP}, 
and for the J and H bands we used the Keck\_NIRC2.J and Keck\_NIRC2.H filter functions
available from SVO Filter Profile Service\footnote{http://svo2.cab.inta-csic.es/theory/fps/} 
\citep{Rodrigo2012ivoa.rept.1015R,Rodrigo2020sea..confE.182R}.
The above-mentioned PHOENIX spectrum and these filter functions are plotted in Fig. \ref{phx_spectrum_and_filter_function}.
To derive colour indices from the PHOENIX model, the synthetic spectral flux, weighted 
with the filter function, is integrated over each band, and then converted into magnitudes.
For this last step, a theoretical spectral flux reference to a `white' A0 standard 
star, such as Vega, is required. We use the synthetic PHOENIX spectrum with the physical parameters
T$_{eff}$ = 9600 K, log(g) = 4.0, and [Fe/H] = -0.5.
The results of these estimations are shown in Fig. \ref{phx_colour_indices}. 
Regarding the colour indices, in general the colour index increases with
decreasing effective temperature; however, as demonstrated in Fig. \ref{phx_colour_indices}, 
the values for the B-V colour index actually decrease with effective temperature once
this latter drops below 3600 K.

\section{\ion{Ca}{II} H\&K flux estimation}
\label{ca_II_flux_est_discribtion}
To estimate the \ion{Ca}{II} H\&K flux, we used the method described
in \citet{Linsky1979ApJS} and the main points of this method are given
in Sect. \ref{ca_II_hk_flux_est}. Here, we want to add and discuss a few added details.

\subsection{Spectrophotometry: counts of a 50 \AA\  continuum window vs. airmass}
\begin{figure}
\centering
\includegraphics[scale=0.55]{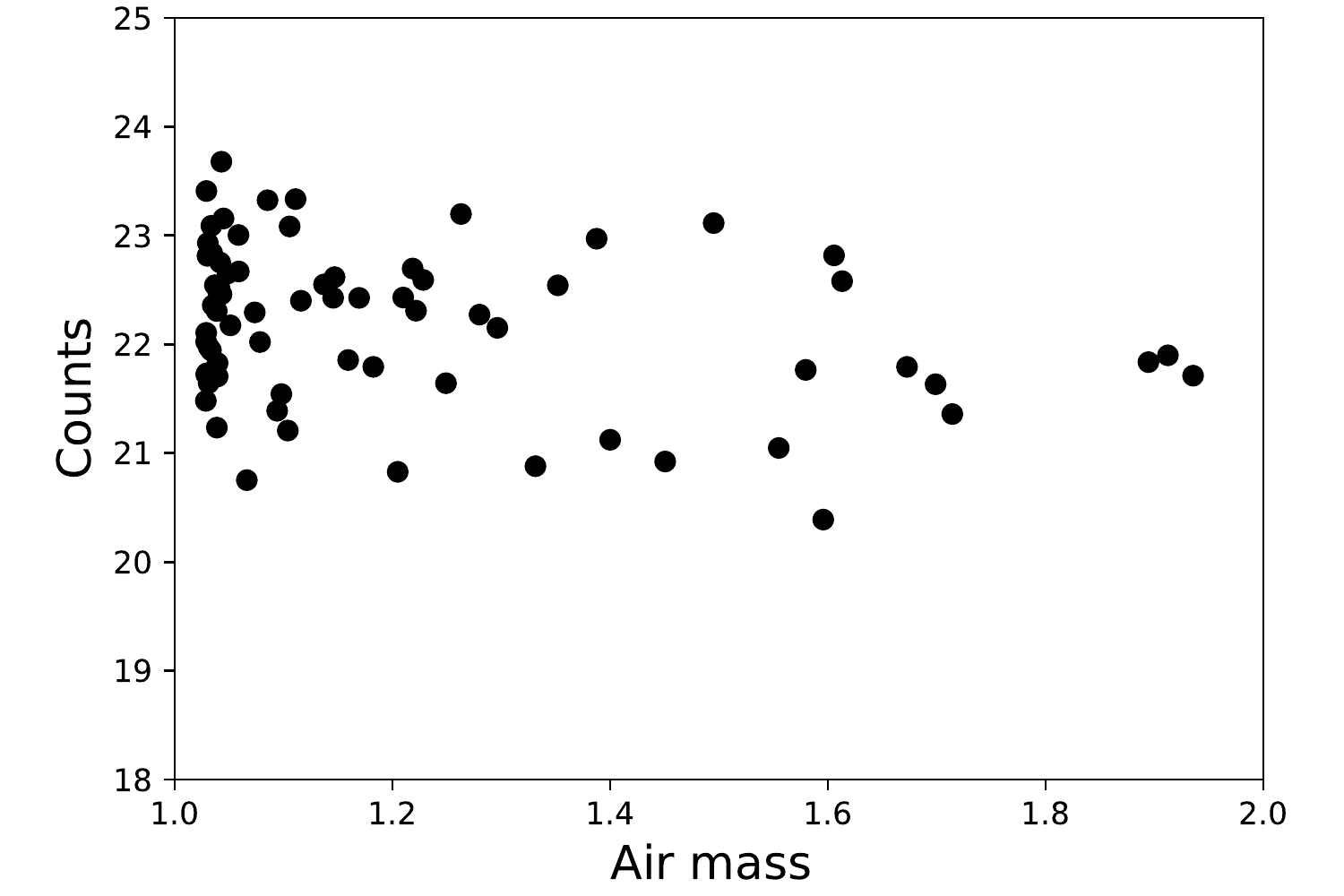}
\caption{Continuum counts without the counts in the 3 \AA\  bandpass of \ion{Ca}{II} H\&K lines vs. air mass.}
\label{check_airmass}
\end{figure}

For the \ion{Ca}{II} H\&K flux estimation with the method described in \citet{Linsky1979ApJS},
a spectrophotometric standard star is required to create an instrumental response 
function. Unfortunately, the air mass of the target star and the observed
spectrophotometric standard star can be different, and so can the extinction, which
then causes a distortion or tilt of the derived spectral energy distribution (SED)
of each flux-corrected spectrum. To test the impact of this problem, which should become 
worse at large airmass, we integrated the counts in the 50 \AA\  bandpass window centred 
at 3950 \AA~(where the counts in the 3 \AA~ bandpass of \ion{Ca}{II} H\&K lines 
are not considered). These continuum counts are plotted here against air mass; 
see Fig. \ref{check_airmass}. Fortunately, we do not find any significant correlation 
between the flux-corrected continuum counts
and the air mass, which suggests that this possible problem is sufficiently well 
controlled by our procedures. 

\subsection{$V-R$ colour index--T$_{\rm{eff}}$ relation}
In the conversion of the count ratio ($I_{HK}$; see Eq. \ref{linsky_flux_count_ratio}), 
the flux depends on the colour index $V-R$; see Eq. \ref{linsky_const_flux}. However, 
the AAVSO database does not contain $V-R$ values for the same days for which we have 
TIGRE spectra. Furthermore, we did not have enough data to derive  
a relation between effective temperature and $V-R$ value.

\begin{figure}
\centering
\includegraphics[scale=0.55]{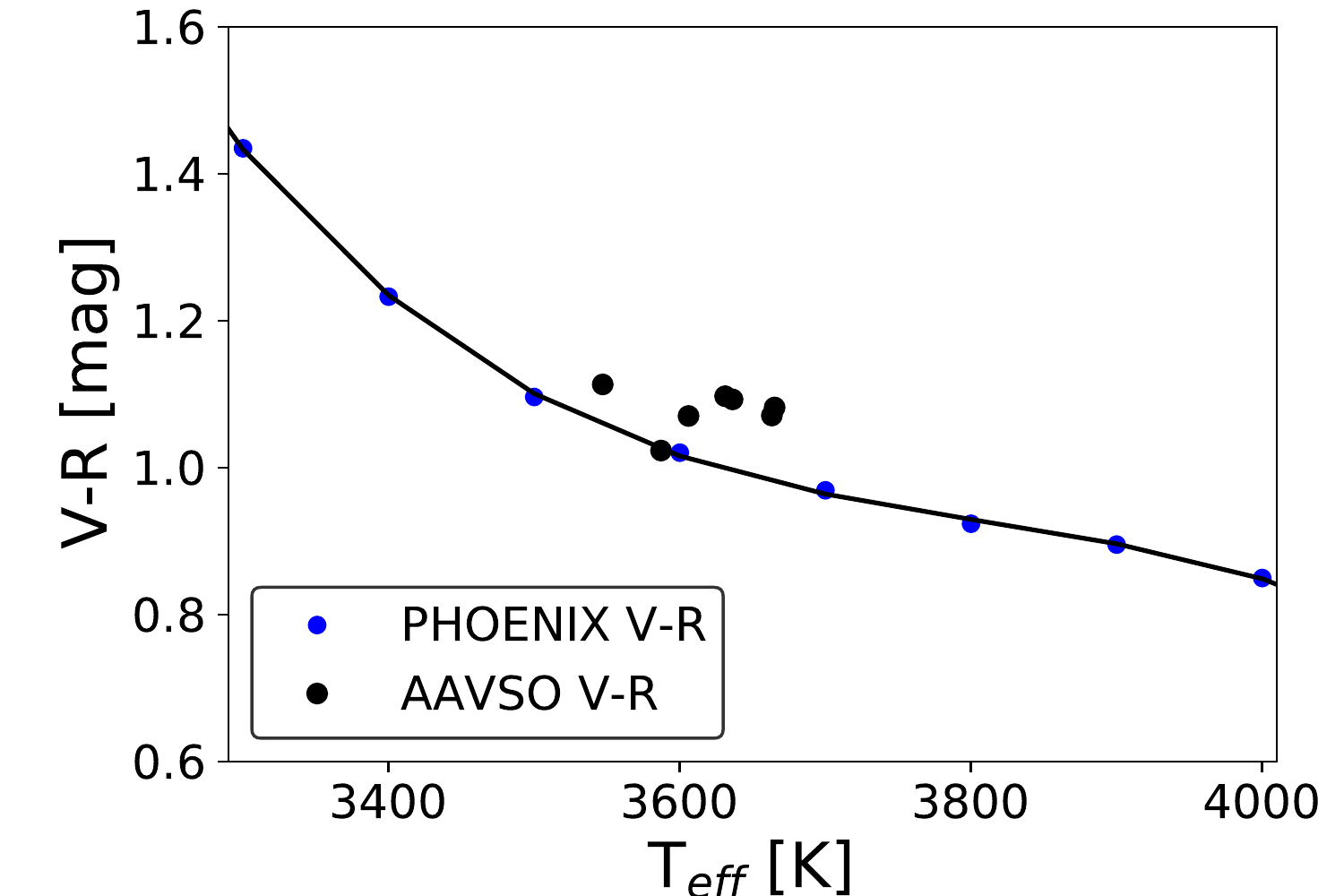}
\caption{PHOENIX $V-R$ vs. effective temperature : Blue points: $V-R$ colours derived from PHOENIX spectra over 
        effective temperature. The solid black line indicates the best-fit relation 
        between the two. Black points: $V-R$ colours from the AASVO database over our 
    derived effective temperatures from different days in the time line.}
\label{VR-teff}
\end{figure}

Therefore, we derived the $V-R$ colour index from PHOENIX spectra of the University 
of G\"ottingen database of \citep{Husser2013A&A} as explained above, using models
of different effective temperatures with [Fe/H]=0.0 and a $\log$(g)= 0.0;
see Sect. \ref{phx_colour}. Figure \ref{VR-teff} shows these $V-R$ colours against the 
respective effective temperature, represented by blue points.

To derive the relation between $V-R$ and the effective temperature, we performed 
a least-square fit through:
\begin{eqnarray}\label{teff_con_vr}  
  V-R & = & 1.4331 - 2.3620\cdot10^{-3} x \nonumber \\
   & & + 4.03\cdot10^{-6} x - 2.64\cdot10^{-9} x, 
\end{eqnarray}
where $x$ is $x = T_{eff}-3300 [K]$; this relation is represented by a solid black 
line in Fig. \ref{VR-teff}.

To test the reality of the PHOENIX spectra $V-R$ values and their dependence on 
effective temperature,
we compare these with the observed $V-R$ colours from the AASVO database, 
coinciding in time with some of the days (only 7) for which we could derive an 
effective temperature. These observed $V-R$ values are shown as black points in Fig. \ref{VR-teff}. 

This comparison shows that the observed $V-R$ values are slightly above their 
synthetic counterparts, but the latter still represent the relative 
changes with effective temperature quite well. We conclude that the $V-R$ values derived 
from PHOENIX models and their
synthetic spectra are slightly smaller than in reality, implying that
the real Ca II H\&K fluxes and flux excesses may be a little larger than stated
here. 

\subsection{Photospheric flux}
In the case of $\alpha$ Ori, it is essential to remove the photospheric flux 
contribution from the \ion{Ca}{II} H\&K line in order 
to study the true stellar activity and prevent this latter from mimicking a false degree
of variation in stellar activity. Here, we give a brief description of how we
derive that photospheric flux contribution, following \citet{Linsky1979ApJS},
who considers its part between the \ion{Ca}{II} K1 and H1 points of the 
\ion{Ca}{II} H\&K lines.
\begin{figure}
\centering
\includegraphics[scale=0.55]{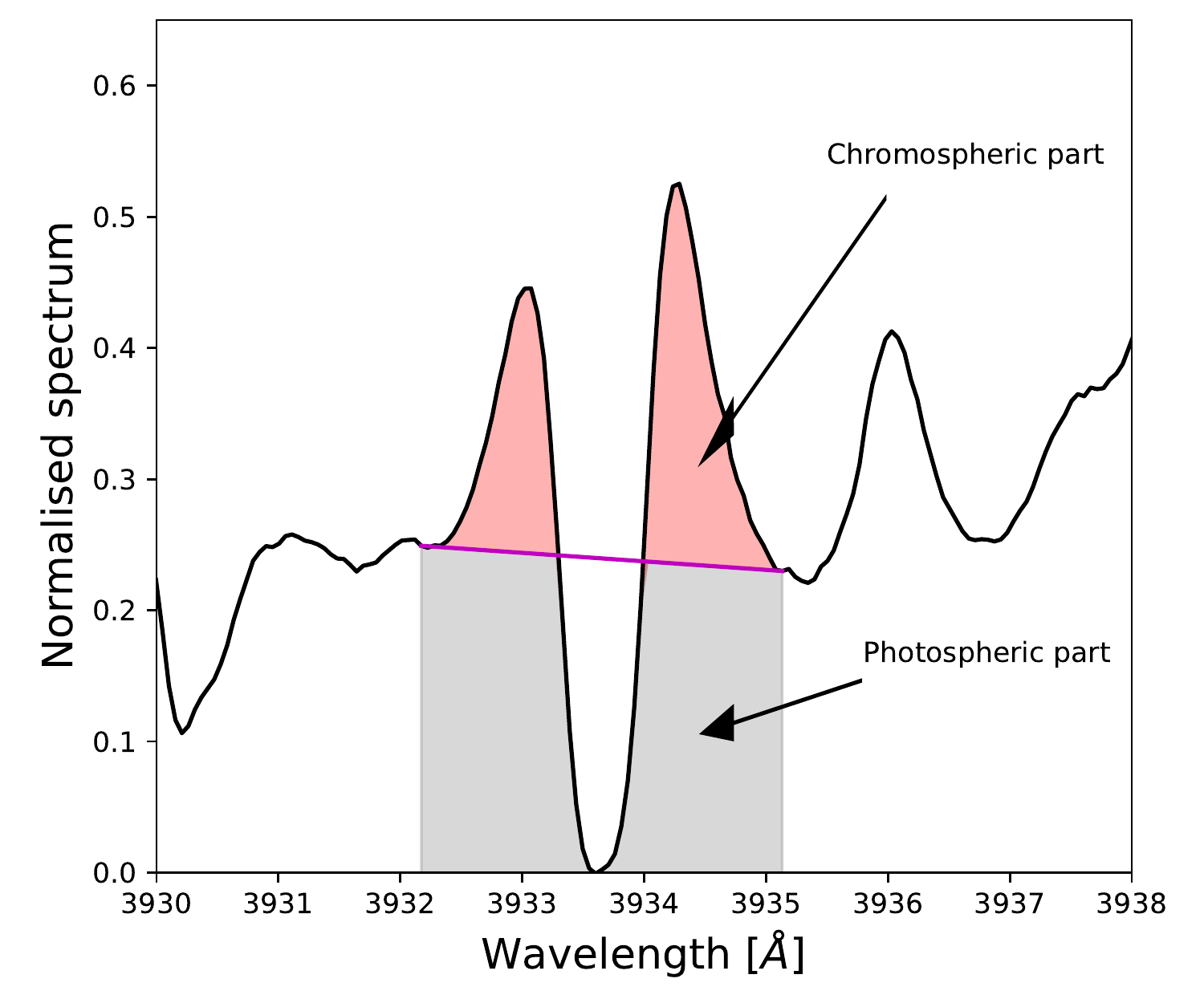}
\caption{\ion{Ca}{II} K line spectrum of $\alpha$ Ori. 
        The magenta solid line shows the fit between the assumed K1V and K1R point.
  The grey area shows the photospheric contribution, and the red area represents the chromospheric emission. }
\label{phot_est}
\end{figure}
To estimate the photospheric flux contribution in the \ion{Ca}{II} H\&K lines in our 
TIGRE spectra, we used the a linear trend between \ion{Ca}{II} K1 and H1 points of the 
\ion{Ca}{II} H\&K lines. 
These are illustrated for the \ion{Ca}{II} K line in Fig. \ref{phot_est}.
Here, the magenta solid line represents the linear trend which splits the line flux into the
the photospheric and chromospheric contribution.   
The grey area in Fig. \ref{phot_est} below the solid magenta line specifies the photospheric line
contribution, and the red area in Fig. \ref{phot_est} above the magenta solid line shows the pure
chromospheric line emission.
\end{appendix}
\end{document}